\documentclass[twocolumn]{aastex61}
\usepackage{longtable}
\submitjournal{ApJ}

\shorttitle{A Calibrator Sample for the M-dwarf Metallicities in APOGEE}
\shortauthors{Souto et al.}

\begin{document}

\title{Stellar Characterization of M-dwarfs from the APOGEE Survey: A Calibrator Sample for the M-dwarf Metallicities}

\correspondingauthor{Diogo Souto}
\email{souto@on.br; diogodusouto@gmail.com}

\author[0000-0002-7883-5425]{Diogo Souto}
\affiliation{Departamento de F\'isica, Universidade Federal de Sergipe, Av. Marechal Rondon, S/N, 49000-000 S\~ao Crist\'ov\~ao, SE, Brazil}
\affiliation{Observat\'orio Nacional/MCTIC, R. Gen. Jos\'e Cristino, 77,  20921-400, Rio de Janeiro, Brazil}

\author[0000-0001-6476-0576]{Katia Cunha}
\affiliation{Observat\'orio Nacional/MCTIC, R. Gen. Jos\'e Cristino, 77,  20921-400, Rio de Janeiro, Brazil}
\affiliation{Steward Observatory, University of Arizona, 933 North Cherry Avenue, Tucson, AZ 85721-0065, USA}

\author{Verne V. Smith}
\affiliation{National Optical Astronomy Observatory, 950 North Cherry Avenue, Tucson, AZ 85719, USA}

\author{C. Allende Prieto}
\affiliation{Instituto de Astrof\'isica de Canarias, E-38205 La Laguna, Tenerife, Spain}
\affiliation{Departamento de Astrof\'isica, Universidad de La Laguna, E-38206 La Laguna, Tenerife, Spain}

\author{Adam Burgasser}
\affiliation{Center for Astrophysics and Space Science, University of California San Diego, La Jolla, CA 92093, USA}

\author{Kevin Covey}
\affiliation{Department of Physics \& Astronomy, Western Washington University, Bellingham, WA, 98225, USA}

\author{D. A. Garc\'ia-Hern\'andez}
\affiliation{Instituto de Astrof\'isica de Canarias, E-38205 La Laguna, Tenerife, Spain}
\affiliation{Departamento de Astrof\'isica, Universidad de La Laguna, E-38206 La Laguna, Tenerife, Spain}

\author[0000-0002-9771-9622]{Jon A. Holtzman}
\affiliation{New Mexico State University, Las Cruces, NM 88003, USA}

\author{Jennifer A. Johnson}
\affiliation{Department of Astronomy, The Ohio State University, Columbus, OH 43210, USA}

\author[0000-0002-4912-8609]{Henrik J\"onsson}
\affiliation{Materials Science and Applied Mathematics, Malm\"o University, SE-205 06 Malm\"o, Sweden}
\affiliation{Lund Observatory, Department of Astronomy and Theoretical Physics, Lund University, Box 43, SE-22100 Lund, Sweden}

\author{Suvrath Mahadevan}
\affiliation{Department of Astronomy \& Astrophysics, Pennsylvania State, 525 Davey Lab, University Park, PA 16802, USA}

\author{Steven R. Majewski}
\affiliation{Department of Astronomy, University of Virginia, Charlottesville, VA 22904-4325, USA}

\author{Thomas Masseron}
\affiliation{Instituto de Astrof\'isica de Canarias, E-38205 La Laguna, Tenerife, Spain}
\affiliation{Departamento de Astrof\'isica, Universidad de La Laguna, E-38206 La Laguna, Tenerife, Spain}

\author{Matthew Shetrone}
\affiliation{University of Texas at Austin, McDonald Observatory, Fort Davis, TX 79734, USA}

\author[0000-0002-0149-1302]{B\'arbara Rojas-Ayala}
\affiliation{Instituto de Alta Investigaci\'on, Universidad de Tarapac\'a, Casilla 7D, Arica, Chile.}

\author{Jennifer Sobeck}
\affiliation{Department of Astronomy, University of Virginia, Charlottesville, VA 22904-4325, USA}

\author[0000-0002-3481-9052]{Keivan G. Stassun}
\affiliation{Department of Physics and Astronomy, Vanderbilt University, 6301 Stevenson Center Ln., Nashville, TN 37235, USA}

\author{Ryan Terrien}
\affiliation{Department of Physics \& Astronomy, Carleton College, Northfield MN, 55057, USA}

\author{Johanna Teske}
\affiliation{The Observatories of the Carnegie Institution for Science, 813 Santa Barbara Street, Pasadena CA, 91101.}
\affiliation{Hubble Fellow}

\author{F\'abio Wanderley}
\affiliation{Observat\'orio Nacional/MCTIC, R. Gen. Jos\'e Cristino, 77,  20921-400, Rio de Janeiro, Brazil}

\author{Olga Zamora}
\affiliation{Instituto de Astrof\'isica de Canarias, E-38205 La Laguna, Tenerife, Spain}
\affiliation{Departamento de Astrof\'isica, Universidad de La Laguna, E-38206 La Laguna, Tenerife, Spain}

\begin{abstract}
We present spectroscopic determinations of the effective temperatures, surface gravities and metallicities for 21 M-dwarfs observed at high-resolution (R $\sim$ 22,500) in the \textit{H}-band as part of the SDSS-IV APOGEE survey. 
The atmospheric parameters and metallicities are derived from spectral syntheses with 1-D LTE plane parallel MARCS models and the APOGEE atomic/molecular line list, together with up-to-date H$_{2}$O and FeH molecular line lists.
Our sample range in $T_{\rm eff}$ from $\sim$ 3200 to 3800K, where eleven stars are in binary systems with a warmer (FGK) primary, while the other 10 M-dwarfs have interferometric radii in the literature. 
We define an $M_{K_{S}}$--Radius calibration based on our M-dwarf radii derived from the detailed analysis of APOGEE spectra and Gaia DR2 distances, as well as a mass-radius relation using the spectroscopically-derived surface gravities.  
A comparison of the derived radii with interferometric values from the literature finds that the spectroscopic radii are slightly offset towards smaller values, with $\Delta$ = -0.01 $\pm$ 0.02 $R{\star}$/$R_{\odot}$.
In addition, the derived M-dwarf masses based upon the radii and surface gravities tend to be slightly smaller (by $\sim$5-10\%) than masses derived for M-dwarf members of eclipsing binary systems for a given stellar radius.
The metallicities derived for the 11 M-dwarfs in binary systems, compared to metallicities obtained for their hotter FGK main-sequence primary stars from the literature, shows excellent agreement, with a mean difference of [Fe/H](M-dwarf - FGK primary) = +0.04 $\pm$ 0.18 dex, confirming the APOGEE metallicity scale derived here for M-dwarfs.
\end{abstract}


\section{Introduction}

M dwarf stars (M-dwarfs) comprise roughly 70\% of all stars in the Milky Way (\citealt{Salpeter1955}, \citealt{Miller1979}, \citealt{Henry2018}). 
Although they represent the most numerous type of star, M-dwarfs remain one of the least studied types of stars in the Galaxy in terms of their chemical abundances. This is primarily due to their complex optical spectra that are blanketed by strong molecular bands such as TiO and VO.
Interest in improving the characterization of M-dwarfs in terms of stellar parameters and metallicities has increased recently as a growing number of Earth-sized exoplanets are increasingly found orbiting M-dwarfs. Small planets are easier to detect around small stars either via radial velocity (RV) or transit methods (\citealt{Charbonneau2007}, \citealt{Gaidos2007}, \citealt{Shields2016}).
Using data from the \textit{Kepler} mission (\citealt{Batalha2013}), \cite{Dressing2015} reported that the occurrence rate of small exoplanets per M-dwarf is 0.56 for Earth-sized planets (1.0--1.5 $R_{\Earth}$) and 0.46 for super-Earths (1.5--2.0 $R_{\Earth}$), see also \cite{Mulders2015}.

Most of the early studies determining metallicities ([Fe/H]) in M-dwarfs were based on photometric calibrations. The pioneering work of \cite{Bonfils2005} determined metallicities for a sample of 20 M-dwarfs in visual binary systems containing a warmer primary of spectral type FGK. Based on the assumption that the secondary M-dwarfs have the same metallicity as their primary star companions, the authors derived a calibration of M-dwarf metallicities as a function of their $M_{K}$ magnitudes and (\textit{V--K}) colors. 
The works of \cite{JohnsonApps2009} and \cite{SchlaufmanLaughlin2010} used similar techniques, also establishing relations between the M-dwarf metallicities and their photometric colors (see also \citealt{Neves2014}, \citealt{Mann2013_binarypaper}, and \citealt{Montes2018}).
Although obtaining M-dwarf metallicities from photometric calibrations is a significant step forward, the internal uncertainties in these measurements (generally of the order of $\sim$0.15---0.20 dex) can be an issue for detailed studies of, for example, possible planet-star connections, or the metallicity distribution of the solar neighborhood. 
Using spectroscopy from both high- and low- resolution spectra to derive metallicities needs to be thoroughly explored, and this can be more easily achieved in the near-IR, since the M-dwarf spectra show shallower and fewer molecular blends in the near-infrared (NIR; J, H, and K bands) than in the optical spectral regions (\citealt{Carmenes2014}, \citealt{Passegger2018}, \citealt{Bean2006}, \citealt{Allard2000}).

\cite{RojasAyala2012} developed a technique to determine metallicities of M-dwarfs from low-resolution ($R\sim$2,000) spectra in the \textit{K}-band 
(2.2 $\micron$), using the equivalent widths (EW) of the Na I doublet lines (2.208 $\micron$ and 2.261 $\micron$), Ca I triplet lines (2.206 -- 2.209 $\micron$) and the H$_{2}$O--K2 index. 
They calibrated a metallicity scale based on 18 M-dwarfs in binary systems with a warmer FGK primary companion. 
The works of \cite{Newton2014}, \cite{Muirhead2014}, \cite{Terrien2015}, and \cite{Mann2013b} used similar techniques to produce spectroscopic M-dwarf metallicity calibrations based on other NIR bands or spectral lines.  
\cite{Veyette2016} argued that metallicity calibrations using Na I and Ca I line equivalent width measurements may present significant uncertainties due to the non-consideration of CO molecular lines, which are an important source of blending in the \textit{K}-band (see \citealt{Tsuji2015}). 
More recently, \cite{Veyette2017} measured equivalent widths of Fe I and Ti I lines from high-resolution \textit{Y}-band spectra of 29 M-dwarfs in binary systems (with a solar-like primary companion) to produce an EW calibration for $T_{\rm eff}$, [Fe/H], and [Ti/Fe], having achieved an internal precision in the derived parameters that is similar to those typically achieved for FGK stars.

Going a step further, the detailed modeling of high-resolution near-infrared spectra of M-dwarfs offers an opportunity to determine precise stellar parameters and metallicities. 
\cite{Onehag2012}, \cite{Lindgren2016} and \cite{Lindgren2017} analyzed high-resolution CRIRES spectra ($R\sim$50,000; \citealt{Kaeufl2004}) of a sample of M-dwarfs showing that their metallicities can be derived from unblended Fe I lines in the \textit{J}-band (1.2 $\micron$) via a spectral synthesis analysis.
In addition, \cite{Lindgren2017} studied the behavior of molecular transitions of FeH, as well as atomic Fe I lines, finding that FeH lines are good indicators of effective temperature in M-dwarfs.
The recent work of \cite{LopezValdivia2019} used the \textit{H}-band IGRINS spectra to derive atmospheric parameters of 254 K-M-dwarf stars matching the spectra with the BT-Settl grids (\citealt{BT-Settl}).

The high-resolution \textit{H}-band spectra from the APOGEE (Apache Point Observatory Galactic Evolution Experiment; \citealt{Majewski2017}) survey constitute a powerful data set to use in determining individual metallicities and detailed chemistry of M-dwarfs.
\cite{Souto2017} derived stellar parameters and metallicities, as well as individual abundances for thirteen elements, in two early-type M-dwarfs ($T_{\rm eff}$ $\sim$ 3850K). 
\cite{Souto2018Ross} extended the same type of spectral analysis to a cooler M-dwarf by deriving stellar parameters and chemical abundances (of eight species) for the mid-spectral type exoplanet--hosting M-dwarf Ross 128 ($T_{\rm eff}$ $\sim$ 3200K).
Using APOGEE spectra as well, \cite{Rajpurohit2018} analyzed 45 M-dwarfs selected from the radial velocity study from \cite{Deshpande2013} and determined atmospheric parameters ($T_{\rm eff}$, log $g$) and metallicities using matches to Phoenix BT-Settl spectral grids. 

In this work, we analyze 21 M-dwarfs with effective temperatures ranging from 3200 to 3950 K and [Fe/H] roughly from -1.00 to +0.25 dex.
One of the main purposes of this study is to compare the metallicity scale of the M-dwarf spectra obtained here from the APOGEE spectra with those obtained from high-resolution optical studies of the warmer primary stars in the literature.
In Section 2 we describe the stellar sample and observations, while the methodology adopted in the derivation of the atmospheric parameters and metallicities is presented in Section 3. Sections 4 and 5 are dedicated to the results and discussion, and we summarize our conclusions in Section 6. 

\section{Observations and Selected sample of M-dwarfs} 

The studied sample is composed of 21 targets: eleven M-dwarfs that are members of wide binary systems containing warmer primaries previously analyzed in the literature using high-resolution spectra (\citealt{Mann2013_binarypaper}, \citealt{Montes2018}), and ten targets that are well-studied field M-dwarfs with interferometric radii measured by \cite{Boyajian2012}. 

The APOGEE spectra analyzed are from the SDSS-IV (Sloan Digital Sky Survey; \citealt{SDSS4}) Data Release 14 (DR14; \citealt{DR14}).
The original APOGEE instrument is a cryogenic high-resolution (R = $\lambda$/$\Delta$$\lambda$$\sim$22,500) multi-fiber (300) \textit{H}-band (1.51 -- 1.69 $\micron$) spectrograph (\citealt{Wilson2010}), operating on the SDSS 2.5-meter telescope (\citealt{Gunn2006}) at Apache Point Observatory. 
A second instrument, virtually identical to the original one is installed in Las Campanas Observatory, but does not concern the data employed in this work.
The targets analyzed here are nearby M-dwarfs having Gaia DR2 distances  $<\sim$ 80 pc (\citealt{Bailer-Jones2018}; Table 1). We note that two stars in our sample (2M11032023+3558117 and 2M11052903+4331357) do not have Gaia DR2 parallaxes and their distances are from \cite{McDonald2017} (also using parallaxes). 
The nearest M-dwarfs in our sample (d $\sim$2.5 -- 7 pc) are quite bright in the \textit{H}-band and these were observed with the APOGEE spectrograph fiber-fed by the 1-m telescope at the APO. 
The APOGEE spectra analyzed here were reduced by the ASPCAP automated pipeline, as discussed in \cite{Nidever2015} and \cite{Holtzman2018}.

\begin{deluxetable*}{lcccccccccccccc}
%\rotate
\tablenum{1}
\tabletypesize{\tiny}
\tablecaption{Stellar Parameters and Metallicities}
\tablewidth{0pt}
\startlongtable
\tablehead{
\colhead{2Mass ID} &
\colhead{J} &
\colhead{H} &
\colhead{Ks} &
\colhead{d(pc)} &
\colhead{$M_{K_{S}}$} &
\colhead{$M_{bol}$} &
\colhead{$R_{\star}$/$R_{\odot}$} &
\colhead{$M_{\star}$/$M_{\odot}$} &
\colhead{$T_{\rm eff}$} &
\colhead{log $g$} &
\colhead{A(C)} &
\colhead{A(O)} &
\colhead{A(Fe)} &
\colhead{[Fe/H]}
}
\startdata
Binaries \\
2M03044335+6144097	&8.877	&8.328	&8.103	&23.5 &6.248	 & 8.884 &0.394	 &0.253	 &3541	&4.65	&8.15 &8.51 &7.19 & -0.26 \\
2M03150093+0103083	&11.622	&11.043	&10.855	&77.5 &6.408	 & 9.029 &0.352	 &0.254	 &3625	&4.75   &7.57 &8.16 &6.54 & -0.91 \\
2M03553688+5214291	&10.885	&10.325	&10.127	&39.6 &7.139	 & 9.800 &0.280	 &0.233	 &3400	&4.91   &8.08 &8.36 &7.00 & -0.45 \\
2M06312373+0036445	&11.077	&10.465	&10.252	&72.1 &5.962	 & 8.564 &0.412	 &0.325	 &3729	&4.72   &7.99 &8.40 &7.06 & -0.39 \\
2M08103429-1348514	&8.276	&7.672	&7.418	&20.9 &5.817	 & 8.458 &0.487	 &0.464	 &3514	&4.73   &8.36 &8.60 &7.51 & +0.06 \\
2M12045611+1728119	&9.793	&9.183	&8.967	&37.6 &6.091	 & 8.756 &0.458	 &0.505	 &3384	&4.82   &8.08 &8.31 &6.93 & -0.52 \\
2M14045583+0157230	&10.129	&9.483	&9.269	&51.8 &5.697	 & 8.319 &0.489	 &0.458	 &3621	&4.72   &8.34 &8.62 &7.60 & +0.15 \\
2M18244689-0620311	&9.659	&9.052	&8.795	&39.6 &5.807	 & 8.473 &0.524	 &0.589	 &3376	&4.77   &8.38 &8.68 &7.66 & +0.21 \\
2M20032651+2952000	&9.554	&9.026	&8.712	&16.0 &7.691	 &10.382 &0.235	 &0.273	 &3245	&5.13   &8.61 &8.87 &7.61 & +0.16 \\
2M02361535+0652191	&7.333	&6.793	&6.574	&7.2  &7.287	 & 9.962 &0.271	 &0.245	 &3331	&4.96   &8.31 &8.60 &7.33 & -0.12 \\
2M05413073+5329239	&6.586	&5.963	&5.759	&12.3 &5.309	 & 7.899 &0.541	 &0.644	 &3791	&4.78   &8.48 &8.73 &7.67 & +0.22\\
\hline                                                                   
Interferometric radii \\                                                  
2M11032023+3558117	&4.203	&3.640	&3.254	&2.6  &6.179	 & 8.809 &0.400	 &0.352	 &3576	&4.78   &8.07 &8.43 &6.99 & -0.46 \\
2M11052903+4331357	&5.538	&5.002	&4.769	&4.8  &6.363	 & 8.992 &0.367	 &0.303	 &3579	&4.79   &8.06 &8.35 &6.87 & -0.58 \\
2M00182256+4401222	&5.252	&4.476	&4.018	&3.6  &6.236	 & 8.877 &0.401	 &0.233	 &3517	&4.60   &8.17 &8.38 &7.02 & -0.43 \\
2M05312734-0340356	&4.999	&4.149	&4.039	&5.7  &5.260	 & 7.848 &0.551	 &0.543	 &3800	&4.69   &8.63 &8.82 &7.80 & +0.35 \\
2M09142298+5241125	&4.889	&3.987	&3.988	&6.3  &4.991	 & 7.571 &0.612	 &0.569	 &3846	&4.62   &8.25 &8.52 &7.62 & +0.17 \\
2M09142485+5241118	&4.779	&4.043	&4.136	&6.3  &5.139	 & 7.722 &0.575	 &0.539	 &3831	&4.65   &8.29 &8.53 &7.71 & +0.26 \\
2M13454354+1453317	&5.181	&4.775	&4.415	&5.4  &5.753	 & 8.370 &0.472	 &0.468	 &3641	&4.76   &8.21 &8.52 &7.21 & -0.24 \\
2M18424666+5937499	&5.189	&4.741	&4.432	&3.5  &6.712	 & 9.348 &0.319	 &0.301	 &3539	&4.91   &8.06 &8.47 &6.97 & -0.48 \\
2M18424688+5937374	&5.721	&5.197	&5.000	&3.5  &7.280	 & 9.947 &0.249	 &0.127	 &3371	&5.00   &8.16 &8.65 &7.00 & -0.45 \\
2M22563497+1633130	&5.36	&4.800	&4.253	&6.9  &5.329	 & 7.911 &0.527	 &0.485  &3831	&4.68   &8.43 &8.67 &7.53 & +0.08 \\
\enddata
\tablenotetext{}{The estimated uncertainties in $T_{\rm eff}$ and log $g$ are 100K, and 0.20 dex, respectively. The mean abundance uncertainties for C, O, and Fe are approximately 0.10 dex.}
\end{deluxetable*}

\section{Abundance Analysis}

The spectral analysis methodology adopted in this study is similar to that presented and discussed in our previous works \cite{Souto2017,Souto2018Ross}. 
The measured spectral lines and the atomic and molecular line lists adopted in the calculations are discussed in \cite{Souto2017,Souto2018Ross}; we used an updated version of the DR14 APOGEE line list (described in \citealt{Shetrone2015} and in V. Smith et al. \textit{in preparation}, and internally labeled as 20150714), which includes the H$_{2}$O line list from \cite{Barber2006} and an FeH line list from \cite{Hargreaves2010}. This is the line list adopted for the SDSS--DR16 (\citealt{DR16}).

We computed spectral syntheses via the semi-automated mode of the BACCHUS wrapper (\citealt{Masseron2016}), which uses the Turbospectrum code (\citealt{AlvarezPLez1998} and \citealt{Plez2012}) and adopted the 1-D LTE (Local thermodynamical equilibrium) plane-parallel MARCS model atmospheres (\citealt{Gustafsson2008}). 
A microturbulent velocity ($\xi$) of 1.00 $\pm$ 0.25 km.s$^{-1}$ was adopted for all stars (see discussion in \citealt{Souto2017}). 
Best fits between the observed and synthetic spectra were obtained via a $\chi$-squared minimization, while we manually fixed the level of the pseudo-continuum of portions of spectra analyzed.
The synthetic spectra were broadened using a Gaussian profile corresponding to the APOGEE spectral resolution (R $\sim$ 22,500), or, a full width at half maximum (FWHM) of $\sim$ 0.73 \AA{}. 
Such resolution imposes a threshold of $\sim$7 km s$^{-1}$ in the detection of the stellar projected rotational velocity, v sin\textit{i}. 
Most of the targets had very low values of v sin\textit{i} that could not be measured, except for two stars that had detectable v sin \textit{i} above this threshold: 2M12045611+1728119, with v sin\textit{i} = 13.5  $\pm$ 2.0 km.s$^{-1}$; and 2M18244689-0620311, with v sin\textit{i} = 10.0 $\pm$ 2.0 km.s$^{-1}$.

%%%%%%%%%%%%%%%%%%%%%%%%%%%%%%%%%%%%%%%%%%%%%%%%%%%%%%
\begin{figure*}
\begin{center}
\includegraphics[angle=0,width=0.95\linewidth]{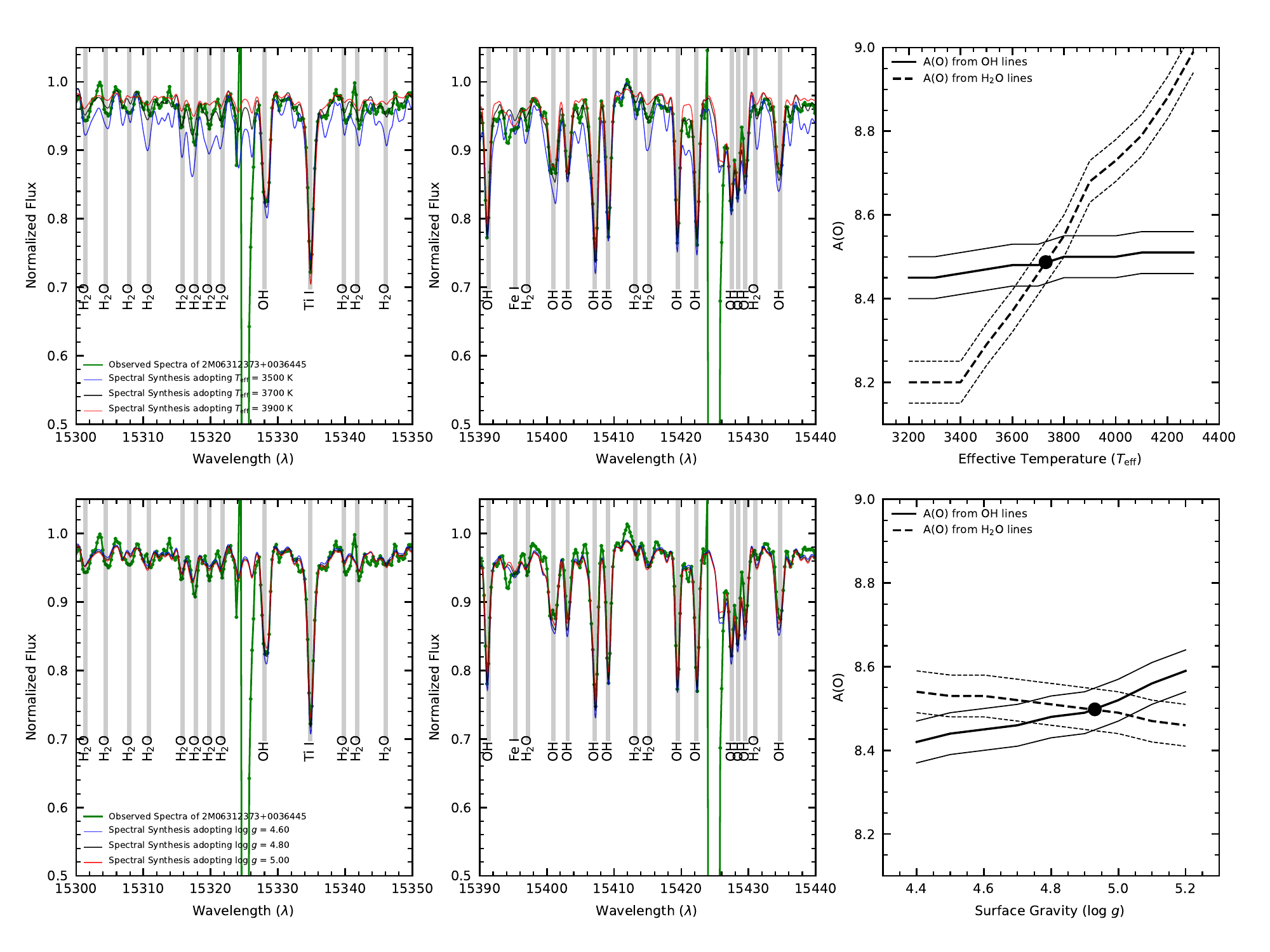}
\caption{
Top and bottom left and middle panels: Portions of the APOGEE spectrum of the M-dwarf 2M06312373+0036445 are shown as green dotted lines. Spectral syntheses computed assuming $T_{\rm eff}$ = 3500, 3700, and 3900 K (log $g$ = 4.8; top panel) and assuming log $g$= 4.6, 4.8 and 5.0 ($T_{\rm eff}$ = 3729K; bottom panel) are shown as blue, black and red solid lines, respectively. 
Right panels: The abundance of oxygen from OH and H$_{2}$O lines as a function of $T_{\rm eff}$ (top) and as a function of log $g$ (bottom). The filled circles represent the pairs $T_{\rm eff}$--A(O) and log $g$--A(O) that indicate an agreement between the abundance indicators.}
\end{center}
\label{Teff_pair}
\end{figure*}
%%%%%%%%%%%%%%%%%%%%%%%%%%%%%%%%%%%%%%%%%%%%%%%%%%%%%%

\subsection{Atmospheric Parameters \& Metallicities}

To estimate the effective temperature of a studied M-dwarf, we derived the oxygen abundances from H$_{2}$O and OH lines for a set of effective temperatures, $T_{\rm eff}$, ranging from 3200--4300 K in steps of 100 K. 
Since the OH and H$_{2}$O lines have different sensitivities to $T_{\rm eff}$ (the OH lines are not very sensitive to $T_{\rm eff}$, while the H$_{2}$O lines are), there is a unique solution for $T_{\rm eff}$ that yields the same oxygen abundance from both OH and H$_{2}$O lines; this is defined by one $T_{\rm eff}$--A(O) pair. 
In this analysis, we initially adopt a log \textit{g} = 4.75.

It should be kept in mind, however, that the derived O abundance is dependent on the C abundance, as expected due to the important role that CO molecules play in the molecular equilibrium pressures. 
For C/O$<$1 (which is expected for all M-dwarfs), C in the stellar atmospheres will tend to be bound in CO, with the remaining O then bound to OH. (This behavior is different for stars with C/O $>$ 1). 
For a given $T_{\rm eff}$, the CO lines in the APOGEE spectra effectively define the carbon abundance of the star, with almost no dependency on the oxygen abundance.
We determine the carbon abundance from fits to the CO lines; we used two CO lines at $\lambda$ = 15978 and 16185 \AA{}, which are well-defined in the spectra of M-dwarfs.  
The derived [C/Fe] and [O/Fe] abundances for the studied M dwarfs, which are in the solar neighborhood, are overall consistent with the expected behavior from chemical evolution. For the most metal-poor star in our sample ([Fe/H] = -0.9), the oxygen abundance is found to be enhanced ([O/Fe]= +0.41).
The C and O abundances will be discussed in full detail in a future paper (D. Souto et al. \textit{in preparation}) where the abundances for eleven other chemical elements will also be presented. 

We present an example of a $T_{\rm eff}$--A(O) diagram in the top panel of Figure \ref{Teff_pair}. The dashed line connects the oxygen abundances from H$_{2}$O lines, while the solid line connects the oxygen abundances from OH lines; changes in $T_{\rm eff}$ by 100 K result in oxygen abundance differences of about $\sim$0.10 dex from H$_{2}$O lines, while we obtain abundance changes less than $\sim$0.02 dex using the OH lines.
The top left and top middle panels of Figure \ref{Teff_pair} show spectral regions dominated by of H$_{2}$O (top-left) and OH (top-middle) lines. The observed spectrum of the sample M-dwarf 2M06312373+0036445 is shown as a dotted green line. 
To illustrate the sensitivity of the H$_{2}$O and OH lines to changes in the adopted $T_{\rm eff}$, we over-plot synthetic spectra for $T_{\rm eff}$ = 3500 (blue line), 3700 (black line), and 3900 K (red line). 
It is clear that the H$_{2}$O lines are the most sensitive to $T_{\rm eff}$ when compared to the other spectral lines in the region, in particular, to the OH lines.

Spectroscopic log $g$s are obtained from the log $g$--A(O) pair. 
The method used to determine the log $g$s is similar to the one adopted for the effective temperature. 
However, instead of changing the $T_{\rm eff}$ we now vary the log $g$ (we test log $g$s from 4.4 to 5.2 dex in steps of 0.10 dex), deriving oxygen abundances with the log $g$s that produce consistent solutions.
In the bottom panel of Figure \ref{Teff_pair}, we present the log $g$--A(O) pair in the same format as in top panel.
The OH lines are more sensitive to changes in log $g$ than $T_{\rm eff}$, while the oxygen abundances derived from the H$_{2}$O lines now decrease as log $g$ increases. 
This method is limited in $T_{\rm eff}$ as the H$_{2}$O lines become very weak in hotter M-dwarfs, with an effective temperature above $T_{\rm eff}$ $\sim$ 3950K and, therefore, this methodology cannot be applied to such warm dwarfs.

%%%%%%%%%%%%%%%%%%%%%%%%%%%%%%%%%%%%%%%%%%%%%%%%%%%%%%
\begin{figure}
\begin{center}
\includegraphics[angle=0,width=1\linewidth,clip]{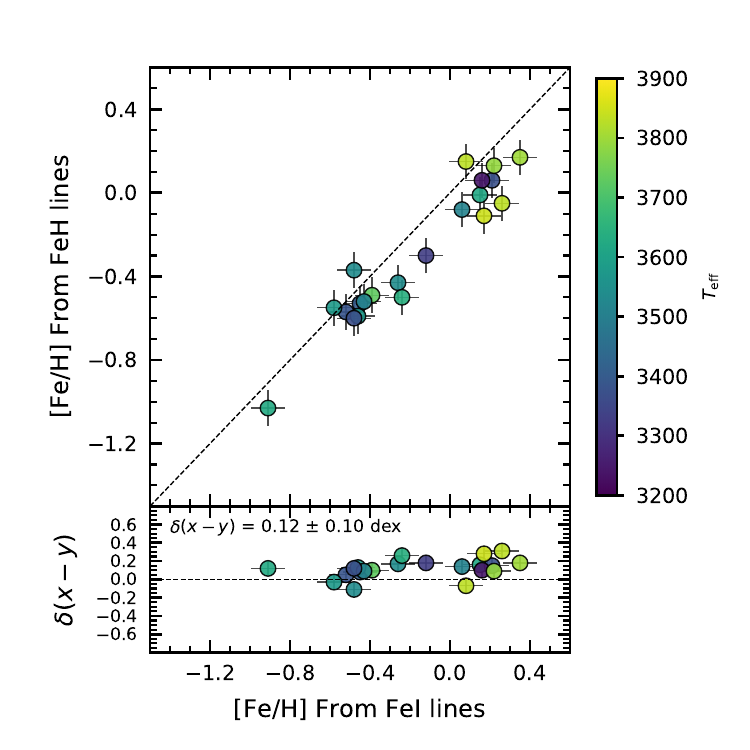}
\caption{The comparison between the iron abundances obtained from the FeH molecular lines and the atomic Fe I lines. The results are color coded by the effective temperature. The difference between these two metallicity indicators is: $\langle$[FeI/H] - [FeH/H]$\rangle$ = +0.12 $\pm$ 0.10 dex.
The residual (\textit{x--y}) diagram is shown at the bottom  panel.}
\end{center}
\label{FeI_FeH_diag}
\end{figure}
%%%%%%%%%%%%%%%%%%%%%%%%%%%%%%%%%%%%%%%%%%%%%%%%%%%%%%

We note that such sensitivity plots (as shown in Figure \ref{Teff_pair}) can also be constructed using Fe I and FeH lines. In the present study, we also investigated consistent solutions for $T_{\rm eff}$ and log $g$ using four spectroscopic indicators: OH and H$_{2}$O lines along with Fe I and FeH lines. However, the effective temperatures and surface gravities derived using Fe I and FeH exhibited small systematics: hotter effective temperatures when compared to fundamental $T_{\rm eff}$ from angular diameters (on average by $\sim$ 150 K), as well as lower surface gravities for a cool main-sequence star (on average by $\sim$ 0.15 dex) than expected. 
These systematics could result from more considerable uncertainties in the g$f$-values of the FeH transitions in our line list (V. Smith et al. \textit{in preparation}); thus, FeH lines were not considered here in deriving this work stellar parameters. 
Nonetheless, deriving the iron abundances only from adjusting the Fe I line profiles, we still find reasonable consistency with the FeH line abundances at the level of $\sim$ 0.10 -- 0.20 dex.
%Nonetheless, this adopted methodology does not force an agreement between the metallicities derived from FeH and Fe I lines, deriving the iron abundances only from adjusting the Fe I line profiles (same transitions as in \cite{Souto2017}), we still find reasonable consistency with the FeH line abundances at the level of $\sim$ 0.10 -- 0.20 dex.} 

Figure \ref{FeI_FeH_diag} presents a comparison of the iron abundances derived from Fe I and FeH transitions; the bottom panel in this figure displays the residual abundance difference between these two indicators. 
The difference between these two metallicity indicators is: $\langle$[FeI/H] - [FeH/H]$\rangle$ = +0.12 $\pm$ 0.10 dex.
We note that this result is in line with the previous finding from \cite{Souto2017} that obtained a systematic difference of 0.10 -- 0.15 dex in the Fe abundances from Fe I and FeH in two M-dwarfs. 

\subsection{Stellar Radii and Masses}

Radii for the M-dwarfs analyzed here were calculated using the definition of luminosity ($L_{\star}$)

\begin{equation}
R_{\star} = \left(\frac{L_{\star}}{4 \pi \sigma T_{\rm eff}^{4}} \right)^{1/2},
\end{equation}
%$L_{\star}$ is the stellar luminosity and 
\noindent where the effective temperatures used in Equation 1 were those derived here spectroscopically.
Luminosities for 16 of the targets were computed directly from the measured bolometric fluxes at the Earth, $F_{\rm bol}$, presented in \cite{Mann2015}, combined with the accurate distances from \cite{Bailer-Jones2018}, which are based on Gaia DR2 (Table 1), assuming no interstellar extinction.  
Five stars in our sample did not have measured values of $F_{\rm bol}$, so we used their absolute \textit{Ks}-band magnitudes, $M_{\rm K}$, along with \textit{Ks}-band bolometric corrections to derive $M_{\rm bol}$ and then $L_{\star}$ using
\begin{equation}
\it L_{\star} = L_{0}10^{-0.4*M_{bol}}
\end{equation}

We used the recommended values from \cite{Mamajek2015} of $M_{\rm bol}$ = 0.00 corresponding to $L_{0}$ = 3.0128x10$^{35}$ erg-s$^{-1}$, which leads to the solar luminosity of 3.828x10$^{33}$ erg-s$^{-1}$ and $M_{\rm bol}$(Sun) = 4.74. 
\textit{Ks}-band bolometric corrections for these five M-dwarfs were derived by using the 16 stars with directly measured luminosities to determine their individual \textit{Ks}-band bolometric corrections from $BC_{\rm K}$ = $M_{\rm bol}$ - $M_{\rm K}$.  These values of $BC_{\rm K}$ define a tight relation of $BC_{\rm K}$ with our derived $T_{\rm eff}$. 
The five M-dwarfs without measured values of $F_{\rm bol}$ span a range in $T_{\rm eff}$ of 3400K--3800K and over this temperature range, a linear relation of $BC_{\rm K}$ with $T_{\rm eff}$ results in a good fit, with $BC_{\rm K}$ = 3.287 - 1.839x10$^{-4}$$\times$$T_{\rm eff}$.
Differences between the values of $BC_{K}$ compared to the linear fit values result in a mean difference of $\Delta$ = 0.00 $\pm$ 0.03 magnitudes, or $\sim$0.03 in $L_{\star}$, and an error in $R_{\star}$ $\sim$1.5\%.
 
Given the derived radii, stellar masses can then be inferred from the fundamental relation:

\begin{equation}
M = g R^{2}/G .
\end{equation}

\subsection{Estimated Uncertainties}

The uncertainties in the determinations are discussed in detail in previous studies and we refer to the abundance sensitivities presented in Table 4 of \cite{Souto2017} and Table 2 of \cite{Souto2018Ross}.
The latter studies estimate that the typical uncertainties in the iron and oxygen abundances are about $\sim$0.1 dex for a change of 65 K in $T_{\rm eff}$, 0.10 dex in log $g$, 0.2 in [M/H] of the model atmosphere and +0.25 km/s in the microturbulent velocity.
We estimate that the typical uncertainties in the iron and oxygen abundances are about $\sim$0.1 dex for a change of $\sim$100 K in $T_{\rm eff}$, 0.2 dex in log $g$, 0.2 in [M/H] of the model atmosphere and +0.25 km/s in the microturbulent velocity. 

If we assume that the $\delta$(A(OH)-A(H$_{2}$O)) can differ by up $\pm$ 0.10 dex (which is the typical measurement precision) as shown by the uncertainty lines in the diagram of O abundances as function of the $T_{\rm eff}$ and log $g$; Figure \ref{Teff_pair}), we obtain the typical uncertainty in $T_{\rm eff}$ to be $\pm$ 100 K .
Using the same procedure for estimating the uncertainties in log $g$, we obtain the uncertainty to be $\sigma$(log $g$) $\sim$ 0.20 dex. 

To determine the errors in the derived stellar radii we adopt the same procedure as \cite{Cintia2019} and propagate the mean associated errors in the \textit{Ks}-band (0.022 mag; \citealt{2MASS}), distances ($\sim$ 0.15 pc), and $T_{\rm eff}$ ($\sim$ 100 K) (the variables in Equation 1). 
We obtain that the errors in our stellar radii are about 5\% total in $R_{\star}/R_{\odot}$, or $\sim$ 0.023$R_{\star}/R_{\odot}$ 
We neglect errors due to extinction and note that that we did not use any reddening for our targets.

\section{Results}

The atmospheric parameters, metallicities, radii, and masses for our sample of 21 M-dwarfs are presented in Table 1. These results will be used in the future to calibrate the APOGEE automated pipeline (ASPCAP; \citealt{GarciaPerez2016}) to produce improvements in the abundances of the M-dwarfs observed in the APOGEE survey.

The results from this study can be compared to those from previous studies using complementary techniques. 
Such comparisons can provide insights into the uncertainties and possible systematic effects inherent in the various analysis methods, as well as improved understanding of differences with theoretical stellar models. 
Reviewing the results obtained from different quantitative analysis techniques, along with the predictions/results from models, is a useful and particularly timely endeavor, given the increased observational efforts dedicated to M-dwarfs as exoplanet host stars.

%%%%%%%%%%%%%%%%%%%%%%%%%%%%%%%%%%%%%%%%%%%%%%%%%%%%%%
\begin{figure}
\begin{center}
\includegraphics[angle=0,width=0.8\linewidth]{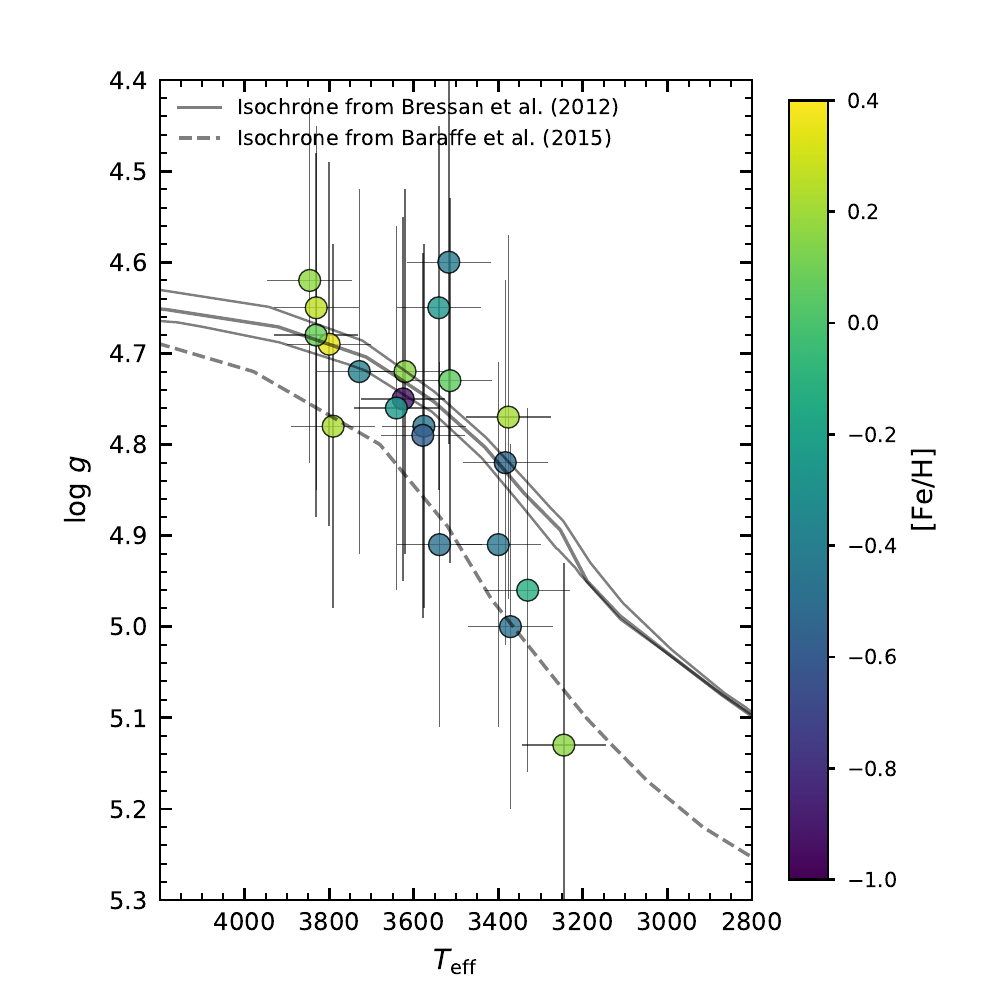}
\includegraphics[angle=0,width=0.8\linewidth]{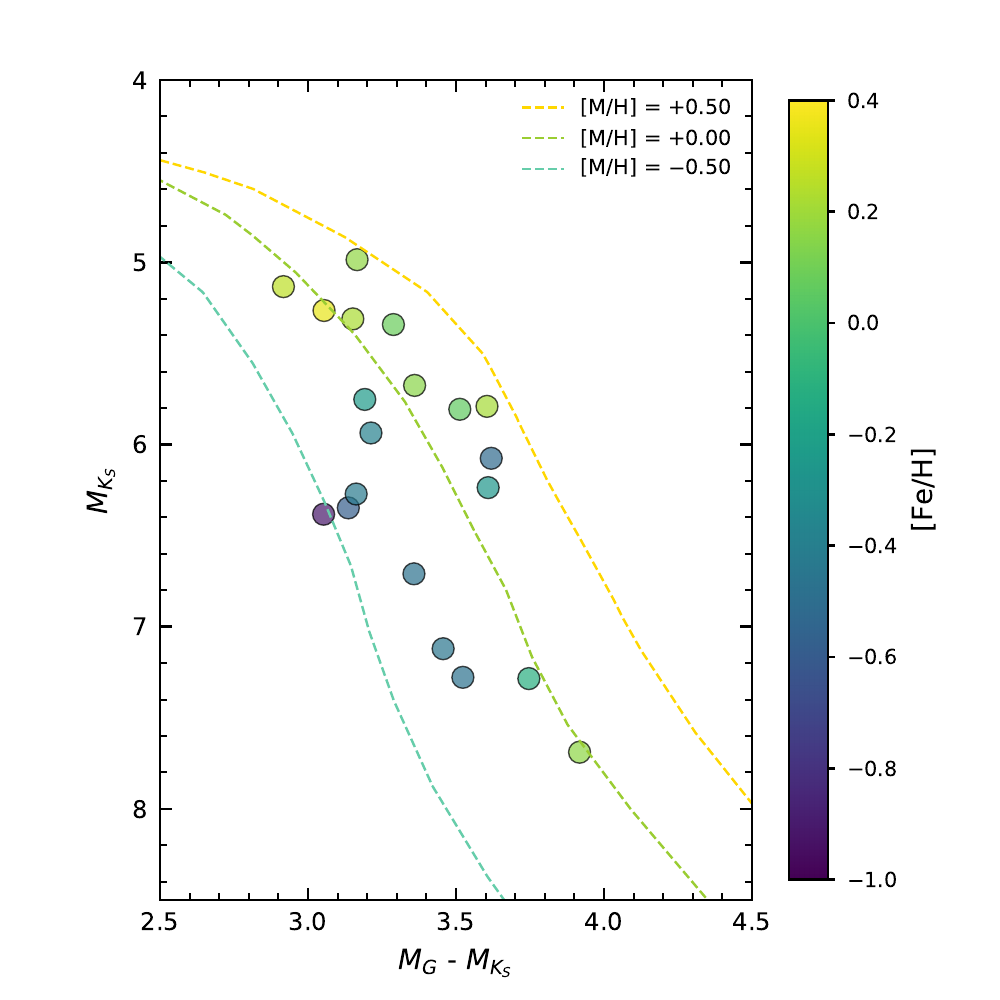}
\caption{Top panel: The Kiel diagram showing the derived stellar parameters for the stars in this study. We also show a solar age--metallicity isochrone (4.5 Gyr and [Fe/H] = 0.00) from \cite{Baraffe2015} and three solar metallicity isochrones from \cite{Bressan2012} corresponding to 1.0, 4.5 and 10.0 Gyr. Bottom Panel: The color-magnitude diagram based on Gaia (\citealt{GaiaCollaborationDR2}) and 2MASS magnitudes. The dashed lined isochrones are from \cite{Bressan2012}. The color bars indicate the derived metallicities for the M-dwarfs.}
\end{center}
\label{Teff_logg_diag}
\end{figure}
%%%%%%%%%%%%%%%%%%%%%%%%%%%%%%%%%%%%%%%%%%%%%%%%%%%%%%

\subsection{Comparisons with Models}

In Figure \ref{Teff_logg_diag} (top panel) we show the Kiel diagram ($T_{\rm eff}$--log $g$) showing the results for stellar parameters in this study. The color bar represents the stellar metallicity of the stars and we also show as a reference the solar age--metallicity isochrone (4.5 Gyr and [Fe/H] = 0.00) from \cite{Baraffe2015} (black--dashed line), as well as three solar metallicity isochrones from \cite{Bressan2012} corresponding to 1.0, 4.5 and 10.0 Gyr (black--solid lines; PARSEC isochrones). 
Overall the derived $T_{\rm eff}$ and log $g$ for the M-dwarfs fall mostly between the two sets of isochrones with some scatter, and perhaps a tendency that the two coolest M-dwarfs in our sample follow more closely the \cite{Baraffe2015} isochrone, while the PARSEC isochrone may better describe the hottest M-dwarfs in our sample. 
In the bottom panel of Figure \ref{Teff_logg_diag}, we show the color-magnitude diagram with absolute magnitudes from Gaia (\citealt{GaiaCollaborationDR2}) and the 2MASS catalogue. For guidance, we also plot three dashed lines representing PARSEC isochrones assuming different metallicities: +0.50 (yellow), 0.00 (green), and -0.50 (blue) dex. The M-dwarfs having near-solar metallicities generally track the solar metallicity isochrone, while the more metal-poor M-dwarfs are displaced and tend to follow the more metal-poor isochrone. 
The overall consistent behavior, within the uncertainties, of our purely spectroscopically derived $T_{\rm eff}$ and log $g$ results when compared to models in the Kiel diagram, as well the comparison of the stellar metallicities and the models in the CMD, reinforces the general consistency between our spectroscopic results and the models. However, a detailed comparison indicates that, as expected, there is room for improvements that could be achieved both on the modeling and on the observational sides.

\subsection{Comparisons with the Literature}

\subsubsection{Effective Temperatures}

%%%%%%%%%%%%%%%%%%%%%%%%%%%%%%%%%%%%%%%%%%%%%%%%%%%%%%
\begin{figure*}
\begin{center}
\includegraphics[angle=0,width=0.45\linewidth]{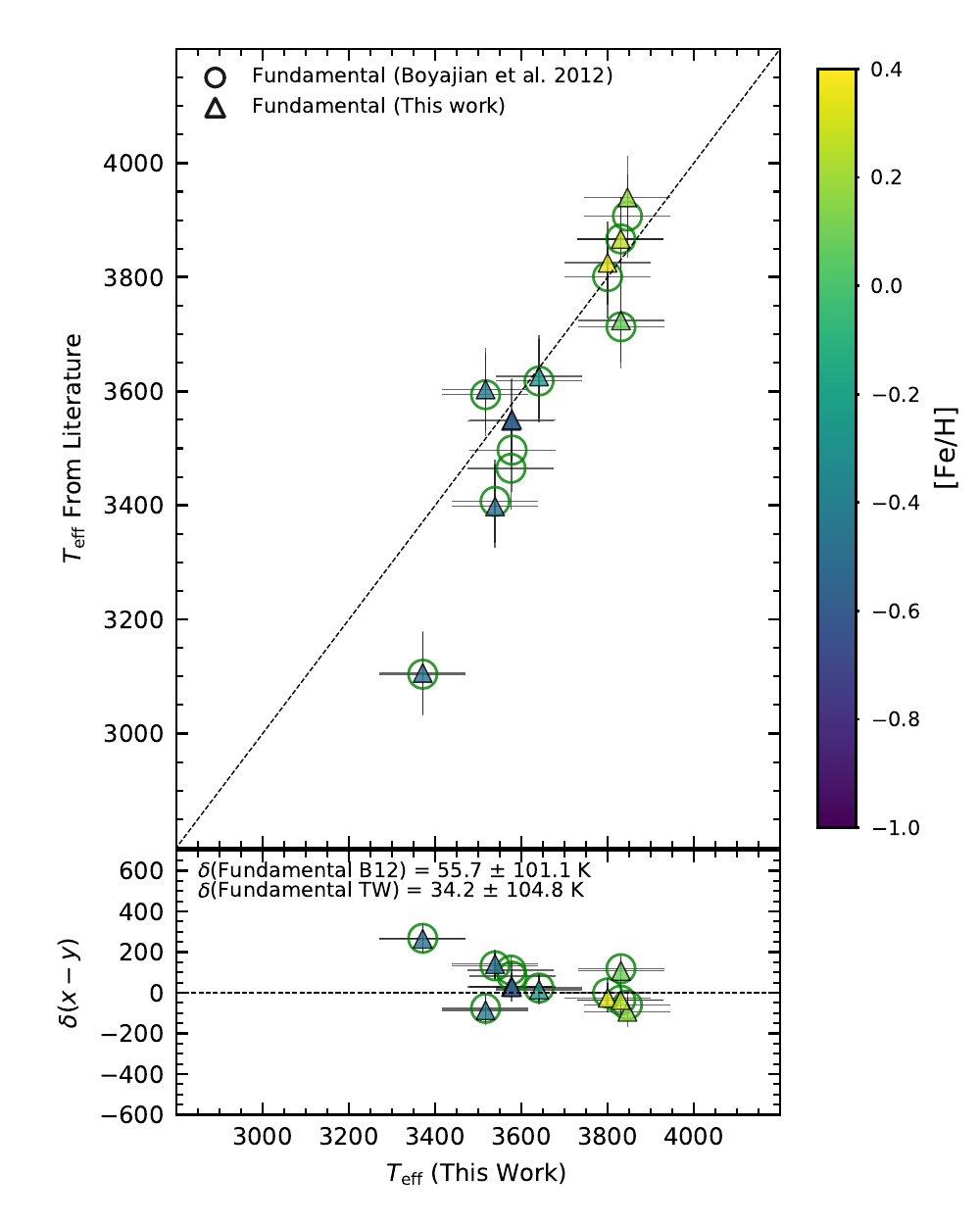}
\includegraphics[angle=0,width=0.45\linewidth]{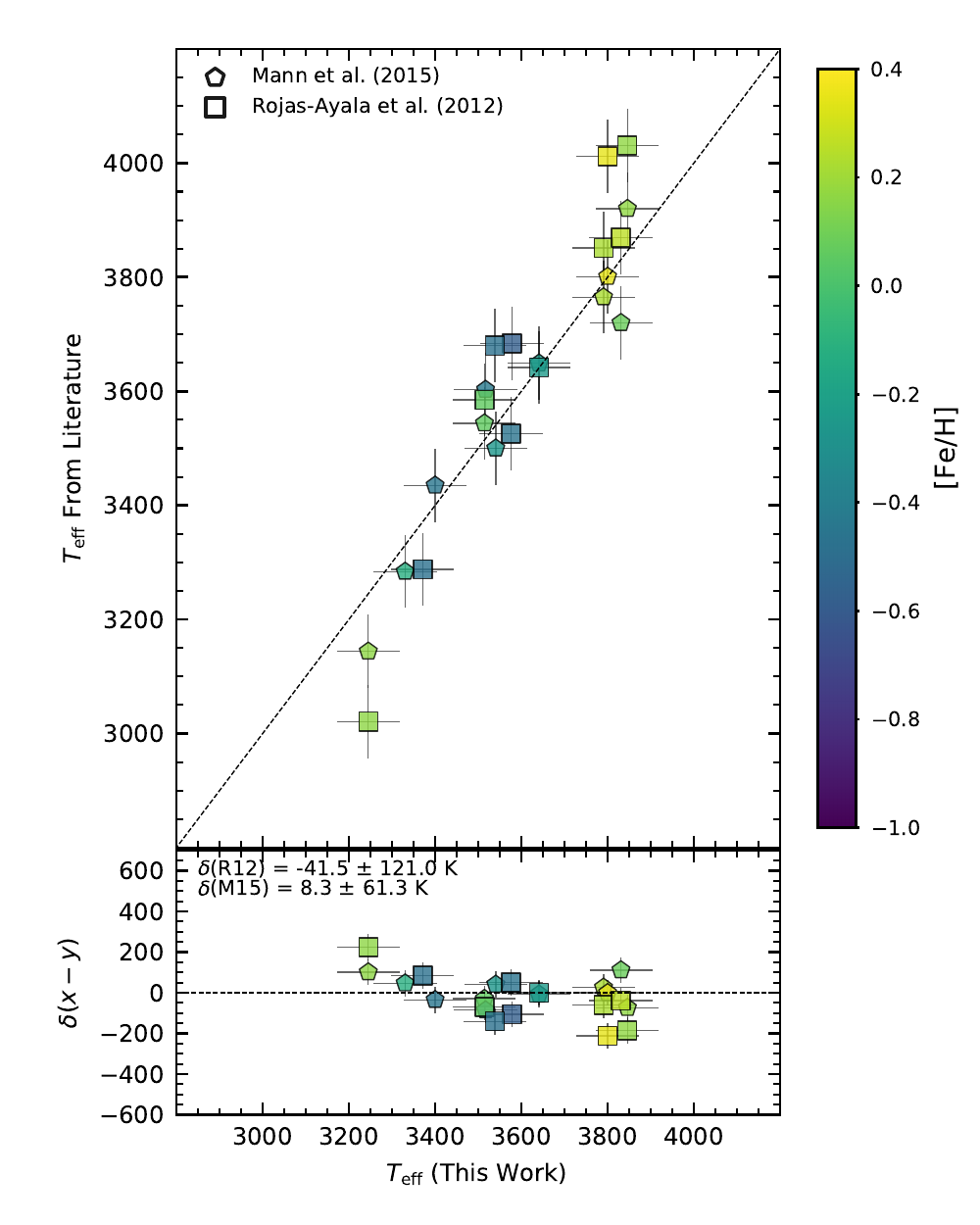}
\caption{
The comparison between the effective temperatures from the APOGEE spectra derived in this study with those from fundamental scale based on measurements of angular diameters (left panel) and other studies in the literature (right panel). Results are color-coded by the metallicity.}
\end{center}
\label{Teff_literature}
\end{figure*}
%%%%%%%%%%%%%%%%%%%%%%%%%%%%%%%%%%%%%%%%%%%%%%%%%%%%%%

The target stars have been well studied in the literature; here we compare our effective temperatures for M-dwarfs in common with the works of \cite{RojasAyala2012},  \cite{Mann2015}, and with the fundamental $T_{\rm eff}$s from \cite{Boyajian2012}.

As mentioned previously, ten target stars are in common with the \cite{Boyajian2012} study, all having fundamental effective temperatures obtained from direct measurements of angular diameters using the interferometric CHARA array.
\cite{Boyajian2012} derived the stellar radius, $R$, using the trigonometric relation $\theta_{\rm LD}$ = 2R/d, where $\theta$ is the measured angular diameter, $d$ is the stellar distance obtained from Hipparcos parallax (\citealt{Hipparcos}) and obtained 
effective temperatures using the flux--luminosity definition: $T_{\rm eff}$ = 2341 ($F_{bol}$/$\theta_{\rm LD}^{2}$)$^{1/4}$, where $F_{bol}$ is in units of 10$^{-8}$ erg-cm$^{-2}$-s$^{-1}$ (where $F_{\rm bol}$ is from \citealt{Boyajian2012}) and $\theta_{\rm LD}$ is in milli-arcseconds.
In this study, we also derived the stellar radii and fundamental effective temperatures using the same angular diameter measurements from \cite{Boyajian2012}, but we now adopt more precise Gaia DR2 distances from \citealt{Bailer-Jones2018} and bolometric fluxes from \cite{Mann2015}. 

%%%%%%%%%%%%%%%%%%%%%%%%%%%%%%%%%%%%%%%%%%%%%%%%%%%%%%
\begin{figure*}
\begin{center}
\includegraphics[angle=0,width=1\linewidth]{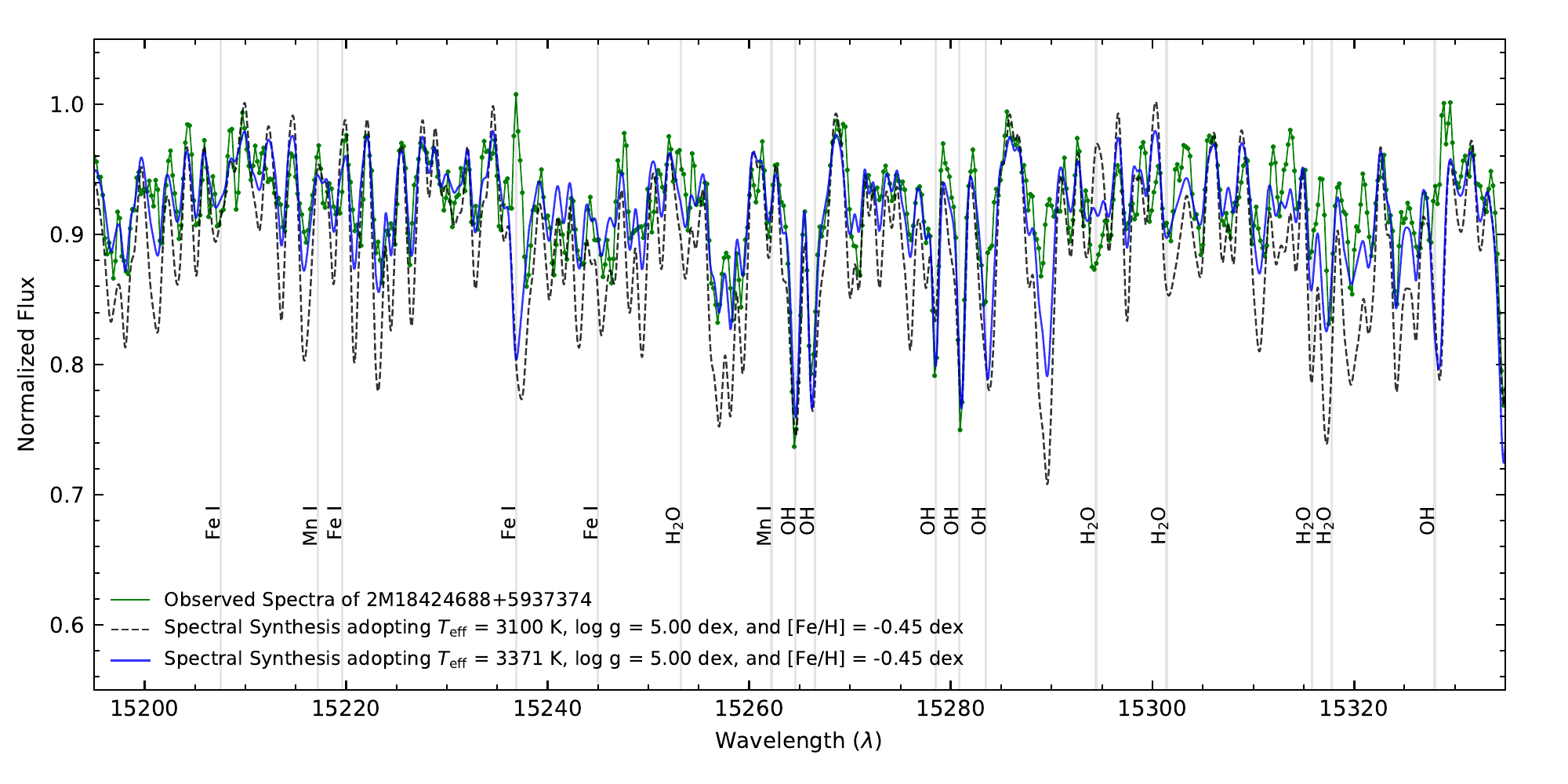}
\caption{A portion of the APOGEE spectra displaying the observed spectrum of 2M18424688+5937374 in green. Two spectral synthesis computed with $T_{\rm eff}$ = 3100 K and 3371 K (in both cases with log $g$ = 5.00 dex and [Fe/H] = -0.45 dex) are shown in dashed black and solid blue lines, respectively.}
\end{center}
\label{spectra_star}
\end{figure*}
%%%%%%%%%%%%%%%%%%%%%%%%%%%%%%%%%%%%%%%%%%%%%%%%%%%%%%

The comparison of the results is presented in Figure \ref{Teff_literature} (left panel), a residual diagram is displayed at the bottom of the Figure and the results are color-coded by the metallicity. The circle and triangle symbols represent the fundamental $T_{\rm eff}$s taken directly from \cite{Boyajian2012} and the fundamental $T_{\rm eff}$s computed in this study using \cite{Boyajian2012} angular diameters, respectively.
There is a small systematic offset in the sense that our spectroscopic $T_{\rm eff}$s are hotter than the fundamental ones:
$\langle$ $T_{\rm eff}$(This work) - $T_{\rm eff}$(fundamental, \citealt{Boyajian2012})$\rangle$ = +56 $\pm$ 101 K, while using the angular measurements together with more precise distances from \citealt{Bailer-Jones2018} we obtain $\langle$ $T_{\rm eff}$(This work) - $T_{\rm eff}$(Fundamental, This work)$\rangle$ = +32 $\pm$ 105 K, indicating a better agreement in the comparison.
We note, however, that one star deviates significantly from perfect agreement, which is also the coolest one in the comparison (2M18424688+5937374; [Fe/H]=-0.45). From the spectroscopic analysis of APOGEE spectra presented here we obtain $T_{\rm eff}$ = 3371 K, while using the angular diameter from \cite{Boyajian2012} and the Gaia DR2 distance we obtain $T_{\rm eff}$ = 3106 K, resulting in a $\delta$ $T_{\rm eff}$ of approximately 250 K. 
In Figure \ref{spectra_star} we show for comparison two synthetic spectra, one corresponding to the stellar parameters obtained here from the APOGEE spectra and another for the $T_{\rm eff}$ obtained d from the \cite{Boyajian2012} angular diameter (adopting log $g$ = 5.00). It is clear that the $T_{\rm eff}$ around 3100 K is too low and does not fit well the APOGEE spectrum for the star 2M18424688+5937374 (shown as dashed black line in Figure \ref{spectra_star}). 
The effective temperature derived in this study of 3371 K is, however, in good agreement with an average of the effective temperatures for this star, $<$$T_{\rm eff}$$>$= 3310 $\pm$ 87  K,
obtained from the literature (\citealt{Mann2015}; \citealt{RojasAyala2012}; \citealt{Valenti1998}; \citealt{Lepine2013}; \citealt{Gaidos2014M_dwarfs}, and \citealt{Gaidos2014}).
It is worth pointing out that interferometric measurements are not completely free from systematic issues, for example, only a fraction of the visibility curve is measured for this star given the small stellar angular size (see Figure 3 in \citealt{Boyajian2012}).
If we remove this star from the comparison, we obtain a $T_{\rm eff}$(This work) - $T_{\rm eff}$(fundamental, This work)$\rangle$ = +8 $\pm$ 75 K, which represents significantly better agreement (or, $\delta$ = -42 $\pm$ 121 K assuming directly the effective temperatures from \citealt{Boyajian2012}).

%%%%%%%%%%%%%%%%%%%%%%%%%%%%%%%%%%%%%%%%%%%%%%%%%%%%%%
\begin{figure*}
\begin{center}
\includegraphics[angle=0,width=1.\linewidth]{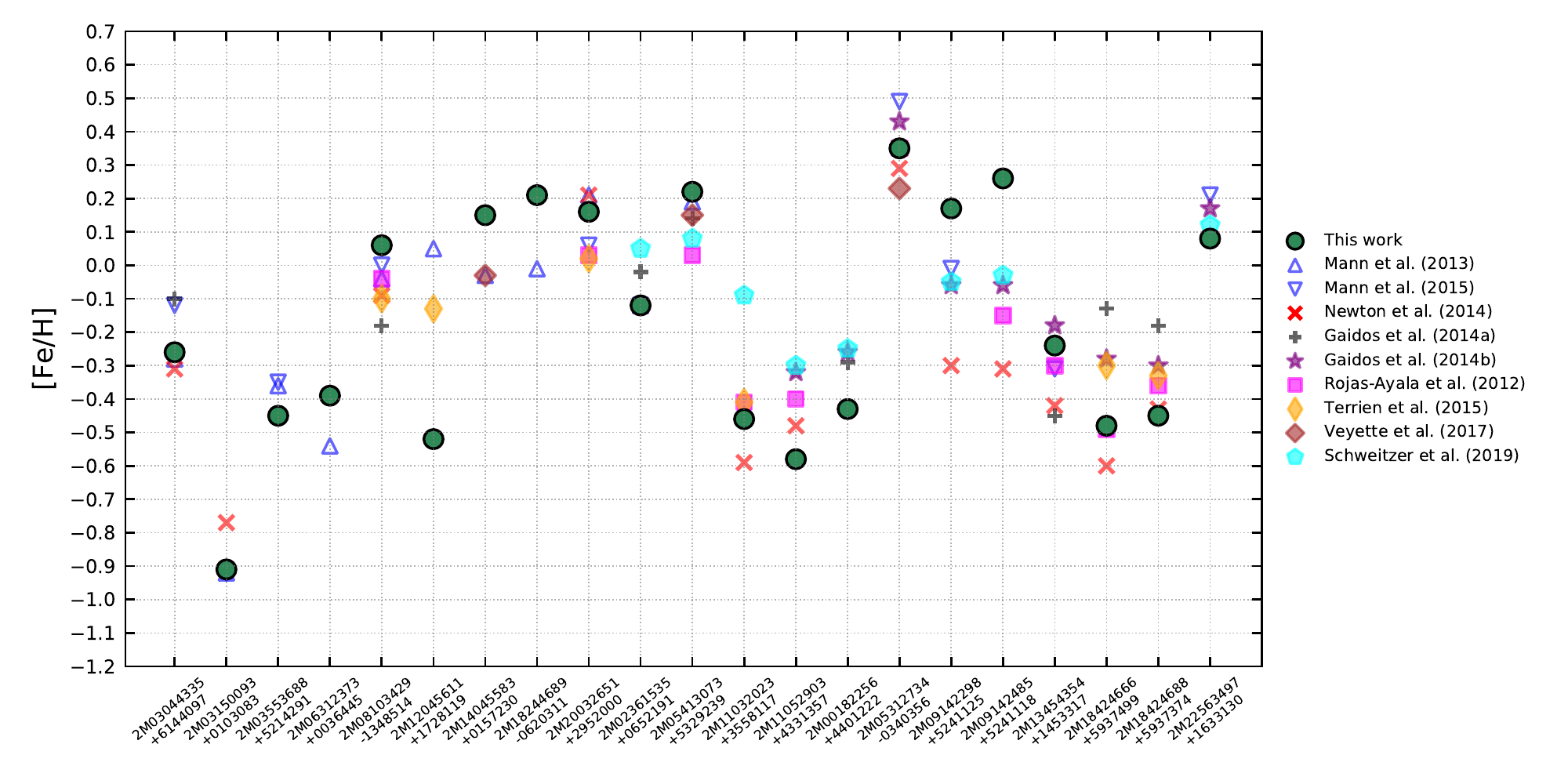}
\includegraphics[angle=0,width=0.5\linewidth]{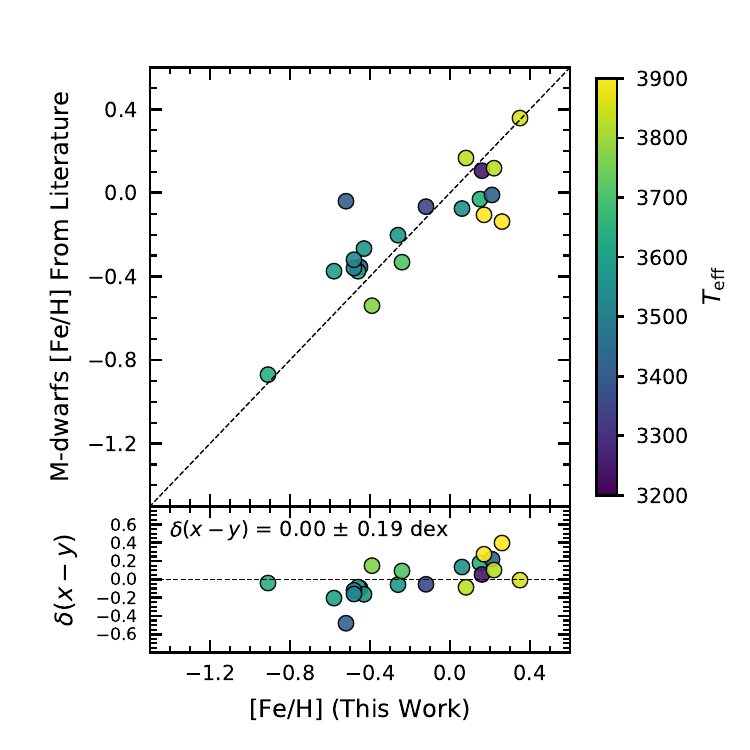}
\caption{Top Panel: The comparison between the metallicities derived in this work from Fe I lines in the APOGEE spectra with results for the same stars from \cite{RojasAyala2012}, \cite{Mann2013_binarypaper}, \cite{Mann2015}, \cite{Newton2014}, \cite{Gaidos2014M_dwarfs}, \cite{Gaidos2014}, \cite{Terrien2015}, \cite{Veyette2017}, and \cite{Schweitzer2019}. Bottom Panel: the average metallicity result for each star compared to this work metallicities. The residual difference is shown in the bottom.}
\end{center}
\label{Met_literature}
\end{figure*}
%%%%%%%%%%%%%%%%%%%%%%%%%%%%%%%%%%%%%%%%%%%%%%%%%%%%%%

\cite{RojasAyala2012} used low-resolution \textit{Ks}-band spectra of a sample of M-dwarfs with effective temperatures derived from the H$_{2}$O--K2 index (\citealt{Covey2010}). \cite{Mann2015} also used low-resolution; however, optical and near-infrared spectroscopy to determine effective temperatures from best matches between their optical spectra and a synthetic grid from BT-Settl Phoenix models (\citealt{BT-Settl}).
Our effective temperatures, which are based upon APOGEE spectra, present reasonably good agreement with the results from the works mentioned above, as can be seen in  \ref{Teff_literature} (right panel, same format as left panel), but there are differences and/or discrepant points both at the low and high effective temperature end and a possible small dependence on the effective temperature, as can be seen from the bottom panel of this figure.
The comparison with results from \cite{Mann2015} shows no offset and a small rms: $\langle$ $T_{\rm eff}$(This work) - $T_{\rm eff}$(\cite{Mann2015}$\rangle$ = +8 $\pm$ 61 K.  
When comparing to \cite{RojasAyala2012}, our $T_{\rm eff}$s are slightly lower with a relatively larger scatter: $\langle$$T_{\rm eff}$(This work) - $T_{\rm eff}$(\cite{RojasAyala2012}$\rangle$ = -29 $\pm$  139 K.

\subsubsection{Stellar Metallicities}

We compiled metallicity results for the studied M-dwarfs from other works in the literature.
In Figure \ref{Met_literature} (top panel) we present the comparison of the derived metallicities with results from the following studies: \cite{RojasAyala2012}, \cite{Mann2013_binarypaper}, \cite{Mann2015}, \cite{Newton2014}, \cite{Gaidos2014M_dwarfs}, \cite{Gaidos2014}, \cite{Terrien2015}, \cite{Veyette2017}, and \cite{Schweitzer2019}. The top panel of Figure \ref{Met_literature} reveals an overall good agreement between the metallicity results, although the Fe abundances derived here for the metal-rich sample ([Fe/H] $>$ 0.00) tend to lie above the abundances derived from the other studies, but not all of them.
We computed an average of the metallicity values available from the literature for each star and a comparison with our results (color-coded by the effective temperature) is shown in the bottom panel of Figure \ref{Met_literature}. 
The mean difference between the metallicities in this case is: 
$\langle$[Fe/H](This work) - $<$[Fe/H](literature)$>$ $\rangle$ = 0.00 $\pm$ 0.19 dex.

\subsubsection{Stellar Radii}

\begin{figure}
\begin{center}
\includegraphics[angle=0,width=1.\linewidth]{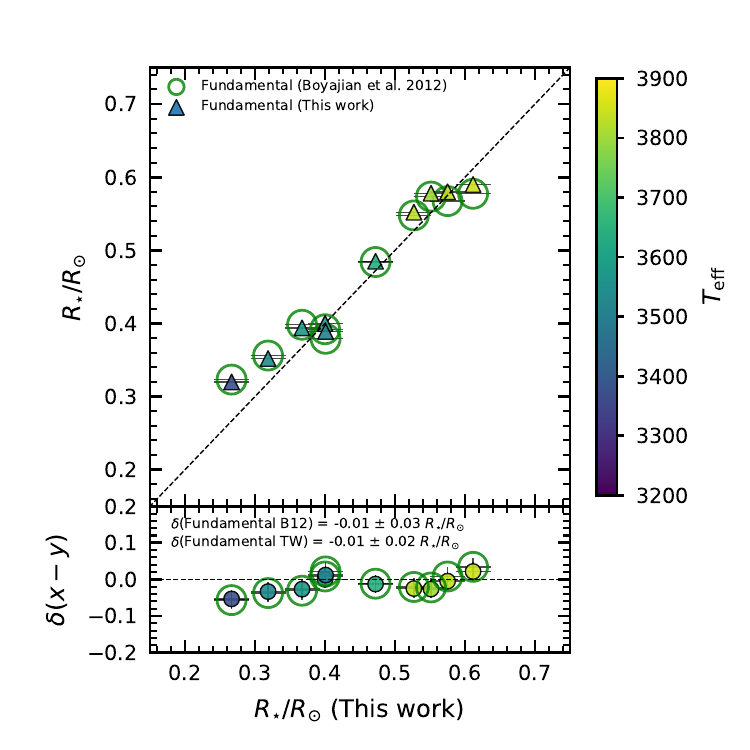}
\caption{The comparison between the derived stellar radii with those measured from interferometry and Hipparcos distances in \cite{Boyajian2012} and those computed in this work using Gaia DR2 distances.} 
\end{center}
\label{radii_radii}
\end{figure}

The comparison of the M-dwarf radii obtained in this work with the interferometric radii reported in \cite{Boyajian2012} and computed in this study using Gaia DR2 distances (Section 4.2.1) is presented in Figure \ref{radii_radii} as a function of the derived effective temperatures (indicated by the color bar). 
There is good agreement between the scales with a slight radius dependence that can be seen in the bottom panel of the figure showing $\delta$ $R_{\star}/R_{\odot}$ versus $R_{\star}/R_{\odot}$. 
As discussed previously for the effective temperature comparison, the slight radius dependence is basically due to one star, or, a single interferometric measurement, notably for the lowest-mass/smallest object in our sample, where only a fraction of the visibility curve is measured because of its small angular size. 
The mean difference between the radii shows just an insignificant offset: $\langle$ $\delta$(This work - \cite{Boyajian2012}) $\rangle$ = -0.01 $\pm$ 0.03 $R_{\star}/R_{\odot}$, while using new distances we obtain $\delta$ = -0.01 $\pm$ 0.02 $R_{\star}/R_{\odot}$.

Stellar radii measured from interferometry are certainly the reference scale in this comparison, for example, the uncertainties in the measured radii in \cite{Boyajian2012} are estimated to be less than about 1 percent (smaller than the symbol size in Figure \ref{radii_radii}), although it is always possible that the measured radii from interferometry may not be completely free from systematics as previously mentioned.
The estimated errors in our derived radii are larger (5 \%) and given the uncertainties we can conclude that there is reasonable consistency between the $R_{\star}$/$R_{\odot}$--scales.

%%%%%%%%%%%%%%%%%%%%%%%%%%%%%%%%%%%%%%%%%%%%%%%%%%%%%%
%%%%%%%%%%%%%%%%%%%%%%%%%%%%%%%%%%%%%%%%%%%%%%%%%%%%%%

\subsection{Comparisons with Photometric Scales for M-dwarfs}

Besides doing direct comparisons for M-dwarfs in common with other studies, it is also of interest to apply photometric calibrations from the literature to estimate effective temperatures for our M-dwarf sample. 
Here we will adopt the photometric $V$-$J$, $r$-$J$, $V$-$H$, and $V$-$K_{s}$ calibrations from \cite{Mann2015} and \cite{Boyajian2012} (discussed in the previous section) and, in addition, the one by \cite{Casagrande2008}, based on a modified version of the InfraRed Flux Method (IRFM; \citealt{Blackwell1979}) with Phoenix model atmospheres.
The stellar $V$, $J$, $H$, $K_{s}$ magnitudes for the stars are taken from the UCAC4 \citep{UCAC4} and 2MASS \citep{2MASS} catalogues, and no reddening correction is considered. 
Overall, the photometric temperature scales are systematically lower than the one derived here from APOGEE spectra; the \cite{Mann2015} $T_{\rm eff}$-scale falls closer to our scale than \cite{Boyajian2012}  or \cite{Casagrande2008}, which also exhibit somewhat larger scatter. 
The mean differences (and standard deviation) are:
$\langle$ $T_{\rm eff}$(This work) - $T_{\rm eff}$(Mann)$\rangle$ =+ 46 $\pm$ 90 K;  
$\langle$ $T_{\rm eff}$(This work) - $T_{\rm eff}$(Boyajian)$\rangle$ =+85 $\pm$ 113 K; and 
$\langle$ $T_{\rm eff}$(This work) - $T_{\rm eff}$(Casagrande)$\rangle$ =+177 $\pm$ 117 K. 
Systematic differences between spectroscopic and photometric $T_{\rm eff}$ were previously reported in, e.g., \cite{Casagrande2008}, \citealt{Onehag2012}, \citealt{Mann2015},  \cite{Schmidt2016}. \cite{Schmidt2016} adopted effective temperatures determined automatically from the APOGEE SDSS-III Data Release 12 (DR12; \citealt{DR12}; \citealt{Eisenstein2011}).
We note that the effective temperatures reported in DR12 are not as accurate in the form as those here, due to the lack of H$_{2}$O and FeH linelists.

\begin{figure*}
\begin{center}
\includegraphics[angle=0,width=0.85\linewidth]{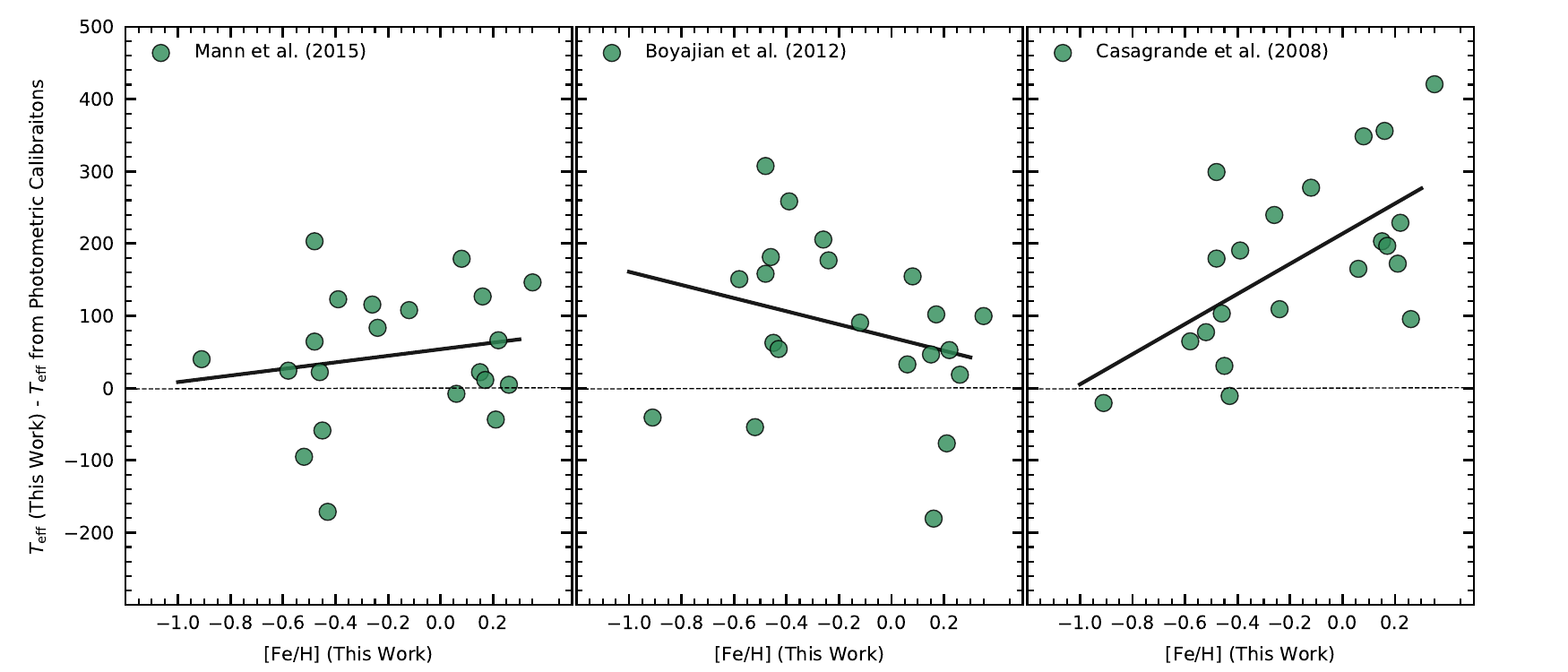}
\caption{The differences ($\delta$ $T_{\rm eff}$) between the effective temperatures derived here from APOGEE spectra and those also derived in this study but using photometric calibrations from the literature by \cite{Mann2015} (left panel), \cite{Boyajian2012} (middle panel) and \cite{Casagrande2008} (right panel). The photometric results correspond to the average $T_{\rm eff}$ obtained from the colors $V$-$J$, $r$-$J$, $V$-$H$, and $V$-$K_{s}$ (when available). The differences are shown versus the stellar metallicity obtained in this study, indicating systematic trends that are represented by solid lines corresponding to simple linear regressions to $\delta$ $T_{\rm eff}$--[Fe/H].}
\end{center}
\label{metal_teff}
\end{figure*}

Clear trends and significant scatter in the 
$\delta$ $T_{\rm eff}$ (This work -- Photometric) 
as a function of [Fe/H] can be seen in Figure \ref{metal_teff} for the photometric scales by \cite{Mann2015} (left panel), \cite{Boyajian2012} (middle panel) and \cite{Casagrande2008} (right panel); in each case we show (as a black line) a linear regression to $\delta$ $T_{\rm eff}$--[Fe/H].
The $T_{\rm eff}$ differences with the \cite{Mann2015} calibration show the smallest trend as a function of metallicity.
For the \cite{Casagrande2008} calibration, we observe a significant trend, where metal-rich stars display systematically lower $T_{\rm eff}$ than the ones derived in this study, while the opposite trend (although smaller) is observed using \cite{Boyajian2012} calibration. 

%%%%%%%%%%%%%%%%%%%%%%%%%%%%%%%%%%%%%%%%%%%%%%%%%%%%%%
%%%%%%%%%%%%%%%%%%%%%%%%%%%%%%%%%%%%%%%%%%%%%%%%%%%%%%

\section{Discussion}

The quantitative analysis of the APOGEE high-resolution near-infrared spectra presented in Section 3 can be used to derive purely spectroscopic fundamental stellar parameters: $T_{\rm eff}$, log $g$, and [Fe/H]. 
Without the need for photometric relations for $T_{\rm eff}$ or log $g$, our analysis provides an independent method to determine atmospheric parameters and metallicities based on high-resolution spectroscopy, that yields independent stellar radii and masses. These can then be compared to the same quantities derived from other observational techniques or predictions from models, and to the fundamental radii and masses obtained from low-mass M-dwarf eclipsing binary systems. 

\subsection{Fundamental Parameters of M-dwarfs}

\begin{figure}
\begin{center}
\includegraphics[angle=0,width=0.95\linewidth]{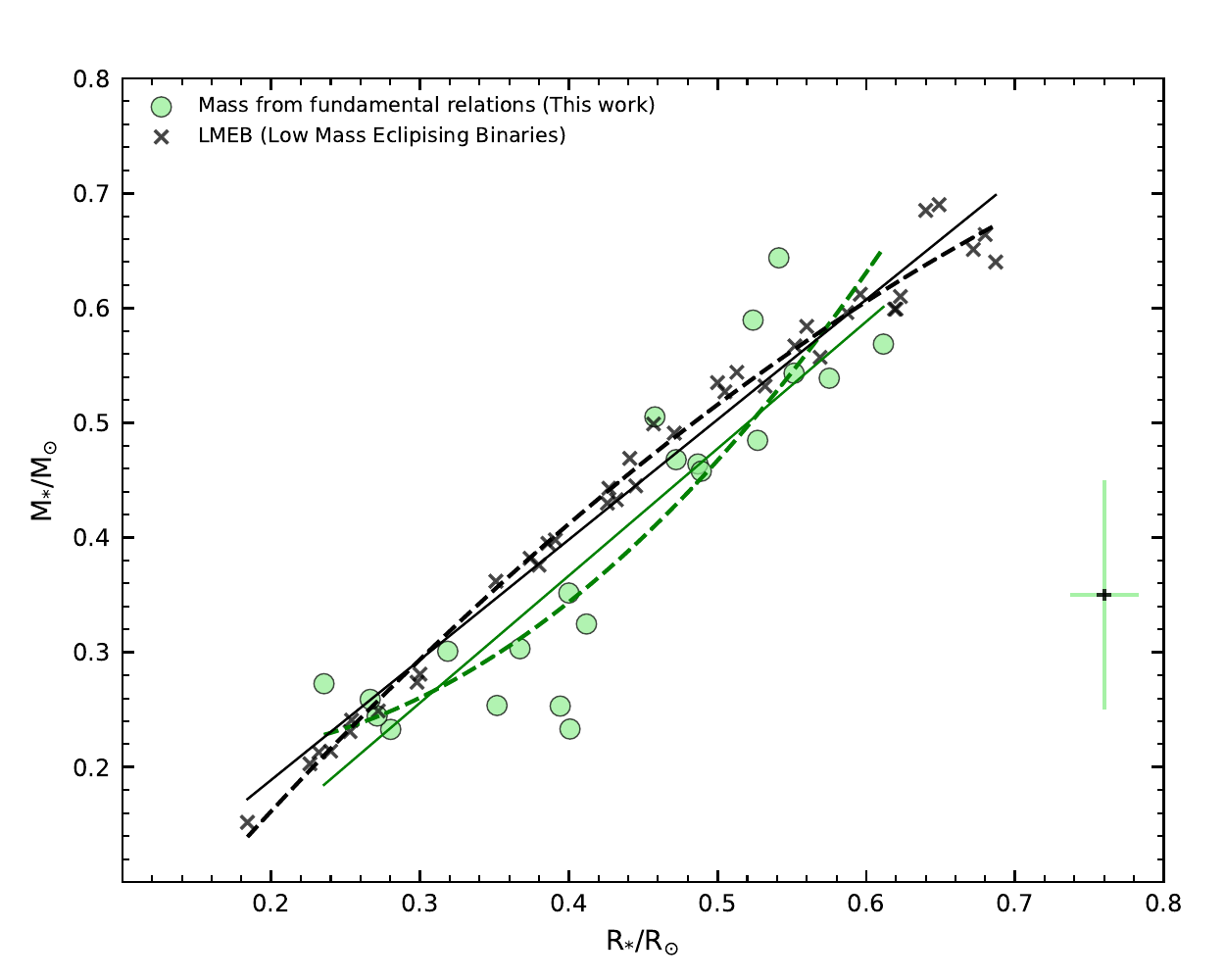}
\caption{The M-dwarf $R_{\star}/R_{\odot}$--$M_{\star}/M_{\odot}$ diagram for the studied stars. The green circle represent this work results and black crosses are from eclipsing binaries M-dwarfs. Linear and quadratic best fits are show as solid and dashed lines (the line colors represent the respective works). The typical error bar for the $R_{\star}/R_{\odot}$ and $M_{\star}/M_{\odot}$ is shown in the bottom right panel of the Figure. }
\end{center}
\label{mass_radii}
\end{figure}

The stellar mass-radius relation obtained in this study is shown in Figure \ref{mass_radii}. 
The derived radii and masses (represented by green circles) are from the spectroscopic results in this study and from fundamental relations (Section 3.2).
Polynomial fits to the data are also shown in the figure as solid and dashed lines, corresponding to 1$^{st}$-degree and 2$^{nd}$-degree polynomial fits, respectively.
The 2$^{nd}$-degree polynomial fit is (green dashed line):

\begin{equation}
    M_{\star}/M_{\odot} = 0.2524 -0.5765(R_{\star}/R_{\odot}) +2.0122(R_{\star}/R_{\odot})^{2}
\end{equation} 

which has an rms = 0.067, while a linear (1$^{st}$-degree) fit is quite similar (solid green line), with a slope and intercept of -0.0760 and 1.107, with an rms of 0.078.  We estimate that the uncertainty in log $g$ dominates the total uncertainty and creates most of the scatter observed in the relation of $M_{\star}$ as a function of $R_{\star}$. We note that the discrepant point that falls off the relation as having too large a mass for its radius is 2M12045611+1728119, which was discussed in Section 3 as being one of two stars in this sample with a measurable rotational velocity, with vsini=13.5 km s$^{-1}$ (the largest vsini in this sample). 

A ``gold standard'' against which to compare the mass-radius relation derived here is the mass-radius relation deduced from the analysis of low-mass M-dwarf eclipsing binary systems; results from such systems (from \citealt{Ribas2003}, \citealt{LopezMorales2005}, \citealt{Torres_Ribas2002}, \citealt{Morales2009a}, \citealt{Morales2009b}, \citealt{Irwin2009}, \citealt{Carter2011}, \citealt{Doyle2011}, \citealt{Irwin2011}, \citealt{Kraus2011}, \citealt{Bass2012}, \citealt{Helminiak2012}, \citealt{Orosz2012a}, \citealt{Orosz2012b}, \citealt{Irwin2018}, \citealt{Torres2018}, and \citealt{Iglesias-Marzoa2019}) are also shown in Figure \ref{mass_radii} as black crosses, along with 1$^{st}$- and 2$^{nd}$-degree polynomial fits to these data.  The mass-radius relation for the eclipsing binary stars exhibits very small scatter. 
The relation for the eclipsing binaries (shown as black solid and dashed curves, respectively) have the following coefficients: $a_{0}$= -0.0203, $a_{1}$ = 1.046, and an rms= 0.018 for a 1$^{st}$-degree fit, and $a_{0}$ = -0.14491, $a_{1}$ 1.67400, and $a_{2}$ = -0.70526, with rms=0.011 for a 2$^{nd}$-degree fit.

The APOGEE spectroscopic results for mass-radius can be compared to those derived from the eclipsing binary M-dwarfs and exhibit a small but measurable offset (as can be seen in Figure \ref{mass_radii}), with the spectroscopic results falling below the relation based on dynamical studies of the eclipsing binary stars.  
The polynomial fits can be compared directly and yield differences of $\Delta$($M_{\star}$/$M_{\odot}$) (This study - Eclipsing Binaries) = -0.01 at 0.2$M_{\odot}$ (or 5\%), -0.03 at 0.3$M_{\odot}$ (10\%), -0.04 at 0.4$M_{\odot}$ (10\%), -0.04 at 0.5$M_{\odot}$ (8\%), and -0.04 at 0.6$M_{\odot}$ (7\%).  These differences are not large and our suspicion is that much of them may reside in uncertainties in the spectroscopic derivation of log $g$.  
%This suspicion is strengthened by the exercise of introducing a small offset into our derived values of log $g$ and using these to recompute the spectroscopic masses. A quite small offset of +0.04 in log $g$ results in the spectroscopic mass-radius relation tracking significantly more closely the relation defined by eclipsing binary M-dwarfs. 

%%%%%%%%%%%%%%%%%%%%%%%%%%%%%%%%%%%%%%%%%%%%%%%%%%%%%%
%%%%%%%%%%%%%%%%%%%%%%%%%%%%%%%%%%%%%%%%%%%%%%%%%%%%%%

\begin{figure}
\begin{center}
\includegraphics[angle=0,width=0.95\linewidth]{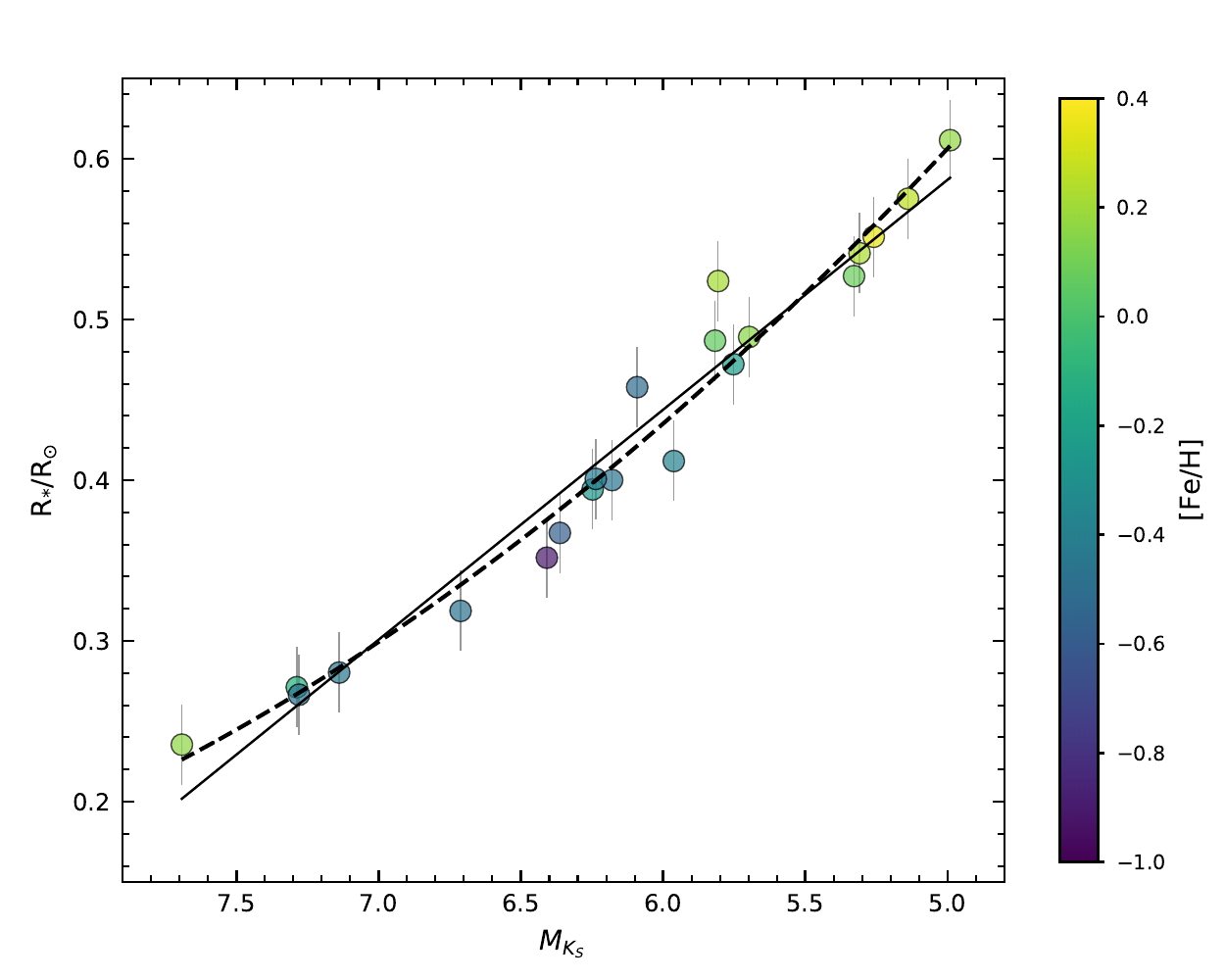}
\caption{The M-dwarf $M_{K_{S}}$--$R_{\star}/R_{\odot}$ diagram for the studied stars. We over-plot two best fits to the data, where the solid and dashed lines represent one and two-degree polynomial fits, respectively.}
\end{center}
\label{radii}
\end{figure}

The relation between the absolute $K$-magnitude, $M_{K_{S}}$, as a function of stellar radius is shown in Figure \ref{radii}.
Also shown are two fits to the data, where the solid and dashed lines represent first- and second-degree polynomials, respectively:

\begin{equation}
    R_{\star}/R_{\odot} = \sum_{n=0}^{n} a_{n} (M_{K_{S}})^{n}
\end{equation}

The coefficients obtained for the 2$^{nd}$-degree polynomial fit to the data are: $a_{0}$ = 1.9932, $a_{1}$ = -0.3659, and $a_{2}$ =  0.0177, (with an rms scatter of 0.008), and for the 1$^{st}$-degree polynomial fit we obtained: $a_{0}$ = 1.3024, and $a_{1}$ = -0.1431, (with an rms scatter of 0.010).
It has been reported previously in the literature that there is a small dependence on metallicity in the $M_{K_{S}}$-radius relation (\citealt{Mann2015}) and that this dependency becomes more important in the metallicity range between [Fe/H] -1.0 and -2.0 (\citealt{Kesseli2019}). 
Our M-dwarf sample covers metallicities between roughly -1 and +0.3 dex and we investigate the metallicity dependency in the $M_{K_{S}}$-radius relation by performing a fit to the data that includes the stellar metallicity as a second independent variable (1 + $b$([Fe/H])).  
A 2$^{nd}$-degree polynomial fit results in coefficients of $a_{0}$ = 1.1621, $a_{1}$ = -0.1069, $a_{2}$ = -0.0020, and $b$ = 0.1182.  We obtain $a_{0}$ = 1.2419, $a_{1}$ = -0.1321, and $b$ = 0.1118, for a 1$^{st}$-degree polynomial best fit.

The inclusion of a metallicity term in the second order polynomial fit results in a small improvement in the residuals (rms = 0.008 and 0.007, for second- and first-order polynomial fits, respectively).
We note that in all cases, the relations are valid for M-dwarfs with $M_{K_{S}}$ in the range of 5.14--7.52. 

As discussed in \cite{Mann2015} the errors in the radius and the absolute magnitudes, $M_{K_{S}}$, are correlated, as these two quantities depend on the adopted distances and their uncertainties.
Propagating the uncertainties into the $M_{K_{S}}$--$R_{\star}/R_{\odot}$ relation results in an internal uncertainty of 0.03$R_{\star}/R_{\odot}$. We adopted the errors in the magnitude to be $\sim$0.022 mag, distances = 0.15 pc, [Fe/H] = 0.10 dex, $T_{\rm eff}$ = 100 K, and in our derived values of $R_{\star}/R_{\odot}$ = 0.023.

We note that the two stars in Figure \ref{radii} having signi\begin{deluxetable}{lccr}
%\rotate
\tablenum{2}
\tabletypesize{\tiny}
\tablecaption{Primary Stars Metallicities}
\tablewidth{0pt}
\startlongtable
\tablehead{
\colhead{2Mass ID} &
\colhead{ID (primary)}&
\colhead{[Fe/H]} &
\colhead{source} 
}
\startdata
Binaries\\
2M03044335+6144097	& HIP 14286&-0.26 $\pm$ 0.05&e,i,j\\ 
2M03150093+0103083	& HIP 15126&-0.85 $\pm$ 0.05&d,h,k,l\\
2M03553688+5214291	& HIP 18366&-0.36 $\pm$ 0.05&d,k\\
2M06312373+0036445	& HIP 31127&-0.54 $\pm$ 0.04&h\\
2M08103429--1348514	& HIP 40035&-0.08 $\pm$ 0.06&d,g,k\\
2M12045611+1728119	& HIP 58919&-0.18 $\pm$ 0.06&b\\
2M14045583+0157230	& HIP 68799&-0.03 $\pm$ 0.04&h\\
2M18244689--0620311	& HIP 90246&-0.18 $\pm$ 0.04&a,d,h\\
2M20032651+2952000	& HIP 98767& 0.21 $\pm$ 0.05&c,d,k,l\\
2M02361535+0652191	& HIP 12114& -0.17 $\pm$ 0.05&f,i,k\\
2M05413073+5329239	& HIP 26779& 0.10 $\pm$ 0.05&j,l\\
\enddata
\tablenotetext{}{Source: (a) \cite{Adibekyan2012}; (b) \cite{Montes2018}; (c) \cite{Bensby2014}; (d) \cite{Carretta2013}; (e) \cite{Galahsurvey2015}; (f) \cite{Ghezzi2010}; (g) \cite{Lambert2004}; (h) \cite{Mann2013_binarypaper}; (i) \cite{Mishenina2008}; (j) \cite{Ramirez2007}; (k) \cite{Ramirez2012}; (l) \cite{Reddy2006}.}
\end{deluxetable}ficantly larger radii at their respective absolute $K$-magnitudes are 2M12045611+1728119 ($M_{K_{S}}$ = 6.076 and $R_{\star}$/$R_{\odot}$ = 0.458), and 2M18244689-0620311 ($M_{K_{S}}$ = 5.791 and $R_{\star}$/$R_{\odot}$ = 0.524), which are the two M-dwarfs in this sample with detectable rotational velocities (\textit{vsin i}= 13.5 and 10.0 km-s$^{-1}$, respectively).
As rotation and magnetic activity are correlated in late-type (FGKM) dwarfs (e.g., \citealt{SuarezMascareo2016}), these two more rapidly rotating M-dwarfs quite possibly are more magnetically active than the other M-dwarfs in this sample.  
It has been suggested that the radii of magnetically active cool, convective dwarf stars (such as M-dwarfs) might be inflated due to magnetic inhibition of convection (e.g. \citealt{Mullan2001} or \citealt{Feiden2014}), or by dark star spots blocking emergent
flux (e.g. \citealt{MacDonald2013}), or a combination of both effects.  For example, \cite{Jackson2018} have recently measured ``inflated'' radii (by about 14\%) in magnetically active Pleiades M-dwarfs.  The larger radii of 2M12045611+1728119 and 2M18244689-0620311 may be related causally to their rapid rotation.

%%%%%%%%%%%%%%%%%%%%%%%%%%%%%%%%%%%%%%%%%%%%%%%%%%%%%%
%%%%%%%%%%%%%%%%%%%%%%%%%%%%%%%%%%%%%%%%%%%%%%%%%%%%%%

\subsection{The Metallicity Scale: M-dwarfs in Binary Systems}

It is reasonable to assume that in a binary system the primary star shares the same (or nearly the same) metallicity and chemical composition as the secondary star. 
The M-dwarf metallicity scale derived here from the APOGEE spectra of eleven M-dwarfs, which are secondary stars in binary systems having warmer companions as primaries, can be checked against the well-defined metallicities of the warmer primaries to search for systematic differences and, ultimately, to serve as an additional validation of the M-dwarf metallicity scale in this study.

We searched the literature for high-resolution spectroscopic studies that previously measured metallicities of the primary stars in our sample. 
Table 2 presents the adopted metallicities for the primaries, which were taken from the following high-resolution works in the literature:
\cite{Adibekyan2012}, \cite{Ammons2006}, \cite{Bensby2014}, \cite{Carretta2013}, \cite{Galahsurvey2015}, \cite{Ghezzi2010}, \cite{Lambert2004}, \cite{Mann2013_binarypaper}, \cite{Mishenina2008}, \cite{Ramirez2007}, \cite{Ramirez2012}, and \cite{Reddy2006}. 
When more than one metallicity value was available, we averaged the results from the different studies, noting that these were quite consistent (typical standard deviations of 0.05 dex or less). 

%%%%%%%%%%%%%%%%%%%%%%%%%%%%%%%%%%%%%%%%%%%%%%%%%%%%%%
\begin{figure}
\begin{center}
  \includegraphics[angle=0,width=1\linewidth,clip]{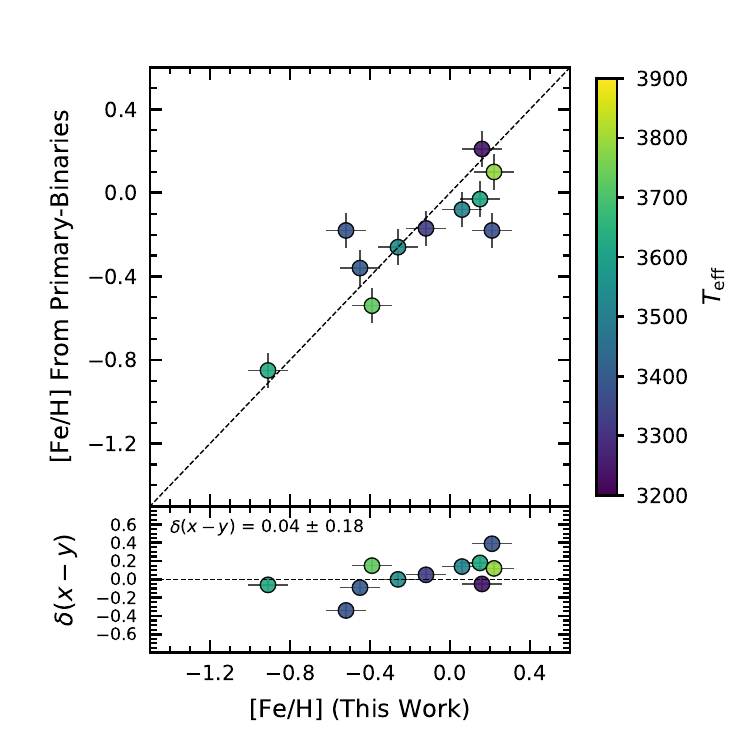}
\caption{The metallicity scale for the M-dwarfs in this study compared with the metallicities compiled from the literature for the primary FGK stars. 
The  diagram is shown at the bottom panel.}
\end{center}
\label{metallicities_primaries}
\end{figure}
%%%%%%%%%%%%%%%%%%%%%%%%%%%%%%%%%%%%%%%%%%%%%%%%%%%%%%

In Figure \ref{metallicities_primaries} we show the comparison between the metallicity scale for the 11 M-dwarfs in binary systems with results obtained from the literature for the warmer primaries (Table 2). Most of the studied stars have metallicities [Fe/H] $>$ -0.8, with the exception of one star that is more metal poor. 
Overall, the agreement between the results is good with a mean difference of $\langle$[Fe/H](This work) - [Fe/H](primaries)$\rangle$ = +0.04 $\pm$ 0.18 dex. 
The systematic offset of +0.04 dex is small (and not statistically significant), in particular when considering the very different effective temperature regime of the two samples (M-dwarfs versus FGK-type stars), the different methodologies and spectral lines analyzed (although Fe I lines are analyzed both in the optical and near-infrared, these are from different excitation potentials, multiplets, etc), as well as the spectral regions analyzed (here APOGEE spectra cover 1.5--1.7 $\mu$m, versus optical spectra for the FGK stars).

The typical uncertainties in our derived iron abundances are about $\sim$0.10 dex (see \citealt{Souto2017,Souto2018Ross}), while the reported uncertainties in the [Fe/H] for the primary stars in the literature are $\sim$ 0.06 dex (not accounting for possible systematic differences in the metallicities for the different adopted works).
Based on these uncertainties we can conclude that the metallicities from the M-dwarfs derived in this study compare well with metallicities from the warmer primary stars within the uncertainties.

\begin{deluxetable}{lccr}
%\rotate
\tablenum{2}
\tabletypesize{\tiny}
\tablecaption{Primary Stars Metallicities}
\tablewidth{0pt}
\startlongtable
\tablehead{
\colhead{2Mass ID} &
\colhead{ID (primary)}&
\colhead{[Fe/H]} &
\colhead{source} 
}
\startdata
Binaries\\
2M03044335+6144097	& HIP 14286&-0.26 $\pm$ 0.05&e,i,j\\ 
2M03150093+0103083	& HIP 15126&-0.85 $\pm$ 0.05&d,h,k,l\\
2M03553688+5214291	& HIP 18366&-0.36 $\pm$ 0.05&d,k\\
2M06312373+0036445	& HIP 31127&-0.54 $\pm$ 0.04&h\\
2M08103429--1348514	& HIP 40035&-0.08 $\pm$ 0.06&d,g,k\\
2M12045611+1728119	& HIP 58919&-0.18 $\pm$ 0.06&b\\
2M14045583+0157230	& HIP 68799&-0.03 $\pm$ 0.04&h\\
2M18244689--0620311	& HIP 90246&-0.18 $\pm$ 0.04&a,d,h\\
2M20032651+2952000	& HIP 98767& 0.21 $\pm$ 0.05&c,d,k,l\\
2M02361535+0652191	& HIP 12114& -0.17 $\pm$ 0.05&f,i,k\\
2M05413073+5329239	& HIP 26779& 0.10 $\pm$ 0.05&j,l\\
\enddata
\tablenotetext{}{Source: (a) \cite{Adibekyan2012}; (b) \cite{Montes2018}; (c) \cite{Bensby2014}; (d) \cite{Carretta2013}; (e) \cite{Galahsurvey2015}; (f) \cite{Ghezzi2010}; (g) \cite{Lambert2004}; (h) \cite{Mann2013_binarypaper}; (i) \cite{Mishenina2008}; (j) \cite{Ramirez2007}; (k) \cite{Ramirez2012}; (l) \cite{Reddy2006}.}
\end{deluxetable}

\section{Conclusions}

We have utilized high-resolution near-IR spectra of 21 M-dwarfs, observed as part of the SDSS IV APOGEE survey, to explore and develop new, purely spectroscopic analysis techniques that can be used to derive fundamental parameters for these cool stars, including atmospheric parameters ($T_{\rm eff}$ and log $g$), as well as metallicities (Fe and O). 
This sample contains eleven secondary M-dwarf stars in binary systems that have hotter FGK main-sequence primaries, plus ten M-dwarfs with interferometric radii measured in the literature.  
Quantitative spectroscopic analysis have been developed which use combinations of H$_{2}$O and OH lines to determine self-consistent values of $T_{\rm eff}$, log $g$, and O abundances via LTE calculations.

We find good agreement, within the uncertainties, in the spectroscopic $T_{\rm effs}$ derived here when compared to results from the literature for stars in common, although the M-dwarf effective temperatures obtained are slightly higher than those derived by photometric relations.

The metallicities of M-dwarfs in binary systems derived here, when compared to those from the literature for the warmer primaries, are in excellent agreement, with no trends with the star's $T_{\rm effs}$.
These results can be used to help calibrate the APOGEE automated ASPCAP (\citealt{GarciaPerez2016}) pipeline to produce improved abundances for the M-dwarfs observed in the APOGEE survey. 

The independent spectroscopic parameters $T_{\rm eff}$ and log $g$ derived in this study can be used, in conjunction with Gaia distances, to calculate the fundamental M-dwarf quantities of radius and mass, which can be compared to radii and masses derived from other techniques. The radii found here via our high-resolution spectroscopic analysis are slightly smaller (by $\sim$0.01 $R_{\star}$/$R_{\odot}$) than the radii inferred from the interferometric observations of \cite{Boyajian2012}.  
Combining the M-dwarf radii determined here with the spectroscopically-derived values of log $g$ results in M-dwarf masses that agree reasonably well with those found via analyses of M-dwarf members of eclipsing binary systems; though there is a systematic offset of $\sim$5-10\%, in the sense that our masses are slightly smaller at a given stellar radius.

The results obtained here, based on a pure, high-resolution spectroscopic analysis of the APOGEE \textit{H}-band spectra of M-dwarfs are important and will help to improve the accuracy of the ASPCAP results for tens of thousands of M-dwarfs, whose scientific importance has increased over the last years due to the discoveries of many ``Earth-like'' exoplanets orbiting some M-dwarf stars.

\acknowledgments

KC and VS acknowledge that their work here is supported, in part, by the National Aeronautics and Space Administration under Grant 16-XRP16\_2-0004, issued through the Astrophysics Division of the Science Mission Directorate. 
SM acknowledge the NSF grant AST-1616636.
DAGH, OZ, and TM acknowledge support from the State Research Agency (AEI) of the Spanish Ministry of Science, Innovation, and Universities (MCIU) and the European Regional Development Fund (FEDER) under grant AYA2017-88254-P.
H.J. acknowledges support from the Crafoord Foundation, Stiftelsen Olle Engkvist Byggm\"astare, and Ruth och Nils-Erik Stenb\"acks stiftelse.

Funding for the Sloan Digital Sky Survey IV has been provided by the Alfred P. Sloan Foundation, the U.S. Department of Energy Office of Science, and the Participating Institutions. SDSS-IV acknowledges
support and resources from the Center for High-Performance Computing at the University of Utah. The SDSS web site is www.sdss.org.

SDSS-IV is managed by the Astrophysical Research consortium for the 
Participating Institutions of the SDSS Collaboration including the 
Brazilian Participation Group, the Carnegie Institution for Science, 
Carnegie Mellon University, the Chilean Participation Group, the French Participation Group, Harvard-Smithsonian Center for Astrophysics, 
Instituto de Astrof\'isica de Canarias, The Johns Hopkins University, 
Kavli Institute for the Physics and Mathematics of the Universe (IPMU) /  
University of Tokyo, Lawrence Berkeley National Laboratory, 
Leibniz Institut f\"ur Astrophysik Potsdam (AIP),  
Max-Planck-Institut f\"ur Astronomie (MPIA Heidelberg), 
Max-Planck-Institut f\"ur Astrophysik (MPA Garching), 
Max-Planck-Institut f\"ur Extraterrestrische Physik (MPE), 
National Astronomical Observatory of China, New Mexico State University, 
New York University, University of Notre Dame, 
Observat\'orio Nacional / MCTI, The Ohio State University, 
Pennsylvania State University, Shanghai Astronomical Observatory, 
United Kingdom Participation Group,
Universidad Nacional Aut\'onoma de M\'exico, University of Arizona, 
University of Colorado Boulder, University of Oxford, University of Portsmouth, 
University of Utah, University of Virginia, University of Washington, University of Wisconsin, 
Vanderbilt University, and Yale University.

\facility {Sloan}

\software{Turbospectrum (\citealt{AlvarezPLez1998}, \citealt{Plez2012}), MARCS (\citealt{Gustafsson2008}), Matplotlib (\citealt{matplotlib}), Numpy (\citealt{numpy}), and Scipy (\citealt{scipy}).}

%\bibliographystyle{yahapj}
%\bibliography{references}

{}

%Tables
\end{document}